\documentclass[times,twocolumn]{aastex63}

\usepackage{CJK}
\usepackage{amsmath, amssymb, bm}
\usepackage{mathrsfs}
\usepackage[nodayofweek]{datetime}

\newdateformat{monthyearday}{%
  \THEYEAR\ \monthname[\THEMONTH] \twodigit{\THEDAY}}
\usepackage{multirow}
\usepackage{lineno}

\makeatletter
\patchcmd{\NAT@citex}
  {\@citea\NAT@hyper@{%
     \NAT@nmfmt{\NAT@nm}%
     \hyper@natlinkbreak{\NAT@aysep\NAT@spacechar}{\@citeb\@extra@b@citeb}%
     \NAT@date}}
  {\@citea\NAT@nmfmt{\NAT@nm}%
   \NAT@aysep\NAT@spacechar\NAT@hyper@{\NAT@date}}{}{}

\patchcmd{\NAT@citex}
  {\@citea\NAT@hyper@{%
     \NAT@nmfmt{\NAT@nm}%
     \hyper@natlinkbreak{\NAT@spacechar\NAT@@open\if*#1*\else#1\NAT@spacechar\fi}%
       {\@citeb\@extra@b@citeb}%
     \NAT@date}}
  {\@citea\NAT@nmfmt{\NAT@nm}%
   \NAT@spacechar\NAT@@open\if*#1*\else#1\NAT@spacechar\fi\NAT@hyper@{\NAT@date}}
  {}{}
\makeatother

\newcommand{\uat}[2]{\href{http://astrothesaurus.org/uat/#1}{#2  (#1)}}

\newcommand{\hi}{H\,{\textsc{\romannumeral 1}}}
\newcommand{\nhi}{{N_{\rm H\,{\textsc{\romannumeral 1}}}}}
\newcommand{\mhi}{{M_{\rm H\,{\textsc{\romannumeral 1}}}}}
\newcommand{\ha}{H{\sc $\alpha$}}
\newcommand{\sfe}{$\rm SFE_{\text{\hi}}$}
\received{2021 December 24}
\revised{2022 May 31}
\accepted{2022 June 13}
\acceptjournal{The Astrophysical Journal}

\shorttitle{On the \hi\ Content of MaNGA Major Merger Pairs}
\shortauthors{Yu et al.}
\graphicspath{{./}{figures/}}
\begin{document}
\begin{CJK*}{UTF8}{gbsn}
\title{On the \hi\ Content of MaNGA Major Merger Pairs}


\author[0000-0003-3230-3981]{Qingzheng Yu (余清正)}
\affiliation{Department of Astronomy, Xiamen University, Xiamen, Fujian 361005, China; \url{fangt@xmu.edu.cn}}

\author[0000-0002-2853-3808]{Taotao Fang (方陶陶)}
\affiliation{Department of Astronomy, Xiamen University, Xiamen, Fujian 361005, China; \url{fangt@xmu.edu.cn}}

\author[0000-0002-9767-9237]{Shuai Feng (冯帅)}
\affiliation{College of Physics, Hebei Normal University, 20 South Erhuan Road, Shijiazhuang, Hebei 050024, China}
\affiliation{Hebei Key Laboratory of Photophysics Research and Application, Shijiazhuang, Hebei 050024,  China}

\author[0000-0002-4781-7056]{Bo Zhang (张博)}
\affiliation{National Astronomical Observatories, Chinese Academy of Sciences (NAOC), Beijing 100101, China}

\author[0000-0002-1588-6700]{C. Kevin Xu (徐聪)}
\affiliation{National Astronomical Observatories, Chinese Academy of Sciences (NAOC), Beijing 100101, China}
\affiliation{Chinese Academy of Sciences South America Center for Astronomy, National Astronomical Observatories, CAS, Beijing 100101, China.}

\author[0000-0001-5751-2347]{Yunting Wang (王允婷)}
\affiliation{Department of Astronomy, Xiamen University, Xiamen, Fujian 361005, China; \url{fangt@xmu.edu.cn}}
\affiliation{Department of Physics and Astronomy, University of British Columbia, 6225 Agricultural Road, Vancouver, V6T 1Z1, Canada.}

\author[0000-0003-2478-9723]{Lei Hao (郝蕾)}
\affiliation{Shanghai Astronomical Observatory, Chinese Academy of Sciences, Shanghai 200030, China}


\begin{abstract}

\noindent The role of \hi\ content in galaxy interactions is still under debate. To study the \hi\ content of galaxy pairs at different merging stages, we compile a sample of 66 major-merger galaxy pairs and 433 control galaxies from the SDSS-IV MaNGA IFU survey. In this study, we adopt kinematic asymmetry as a new effective indicator to describe the merging stage of galaxy pairs. With archival data from the HI-MaNGA survey and new observations from the Five-hundred-meter Aperture Spherical radio Telescope (FAST), we investigate the differences in \hi\ gas fraction ($f_\text{\hi}$), star formation rate (SFR), and \hi\ star formation efficiency (\sfe) between the pair and control samples. Our results suggest that the \hi\ gas fraction of major-merger pairs on average is marginally decreased by $ \sim 15\%$ relative to isolated galaxies, implying mild \hi\ depletion during galaxy interactions. Compared to isolated galaxies, pre-passage paired galaxies have similar $f_{\text{\hi}}$, SFR and \sfe, while pairs during pericentric passage have weakly decreased $f_{\text{\hi}}$ ($ -0.10\pm0.05$ dex), significantly enhanced SFR ($ 0.42\pm0.11$ dex) and \sfe\ ($ 0.48\pm0.12$ dex). When approaching the apocenter, paired galaxies show marginally decreased $f_{\text{\hi}}$ ($ -0.05\pm0.04$ dex), comparable SFR ($ 0.04\pm0.06$ dex) and \sfe\ ($ 0.08\pm0.08$ dex). We propose the marginally detected \hi\ depletion may originate from the gas consumption in fuelling the enhanced $\rm H_2$ reservoir of galaxy pairs. In addition, new FAST observations also reveal an \hi\ absorber ($\nhi \sim 4.7 \times 10^{21} \text{ cm}^{-2}$), which may suggest gas infalling and the triggering of AGN activity.

\end{abstract}


\section{Introduction} \label{sec:intro}
Atomic gas (\hi) plays a significant role in galaxy formation and evolution, fueling molecular gas needed for star formation in the interstellar medium (ISM) and/or cool ionized gas in the circumgalactic medium (CGM) \citep[e.g.,][]{Lehner2011,Borthakur2015,Wang2020}. As a reliable tracer of neutral hydrogen gas, the \hi\ 21 cm emission line has long been used to study the \hi\ content, kinematics, fueling of star formation, and quenching of galaxies in the local universe \citep[e.g.,][]{Meyer2004,Haynes2018,Catinella2018,Wang2020,Zhang2019}.
Also, the associated \hi\ 21 cm absorption line observed in the active galactic nuclei (AGNs) helps explore the gas interactions and co-evolution between AGNs and host galaxies \citep{Gorkom1989,Vermeulen2003,Ger2015}. 

Interactions of galaxy mergers provide an effective way to study the impact of \hi\ gas during galaxy evolution, as recent observations and simulations have connected galaxy mergers with the enhancement of star formation, gas regulation, triggering of AGN, and starburst activities \citep{Mihos1996,DiMatteo2007,Cox2008,Ellison2008,Satyapal2014,Moreno2015,Hani2018}. Although many efforts have been made to explore the gas conditions and star formation during galaxy interactions \citep[e.g.,][]{Hibbard1996,Georgakakis2000,Ellison2015,Zuo2018}, the role of \hi\ gas is still under debate. Using different interacting galaxy samples, some observations of pre- and post-mergers indicate that the \hi\ gas fractions of galaxy mergers are enhanced compared with isolated galaxies \citep{Casasola04,Janowiecki2017,Ellison2018}, while others find no significant difference in \hi\ gas fractions \citep{Ellison2015,Zuo2018} or decreased \hi\ content \citep{Hibbard1996,Georgakakis2000}. Several factors contributed to the confusing results, for example, limited pair sample or lack of robust control sample \citep{Moreno2019}. However, a less explored factor is the lack of rigorously defined merger stages \citep{Pan2019}.

The current definition of merger stage for galaxy pairs has several shortcomings, as the pair selection mainly relies on the projected separation ($d_{\rm p}$) and the difference between line-of-sight velocities  ($\Delta v$) \citep{Ellison2008,Scudder2012,Patton2013,Zuo2018,Zhang2020}. Considering the projection effect, the projected separation ($d_{\rm p}$) may not reveal the physical separation of member galaxies \citep{Soares2007}. Furthermore, the degree of interactions could be different for galaxy pairs with the same separation but at different merging stages\citep{Torrey2012,Moreno2015}. Although some works use morphology as a merger stage indicator \citep[e.g.,][]{Smith2018,Pan2019}, stages for galaxy pairs still lack a quantitative definition. 

With unique data of gas kinematics, recent integral field unit (IFU) surveys of nearby galaxies, such as the Calar Alto Legacy Integral Field Area survey \citep[CALIFA,][]{Sanchez2012} and the Sydney-AAO Multi-object Integral field spectrograph galaxy survey \citep[SAMI,][]{Croom2012} have connected the asymmetry of gas kinematics with galaxy interactions \citep{Barrera2015,Bloom2018}. Using the IFU data from Mapping Nearby Galaxies at APO (MaNGA) survey \citep{Bundy2015}, \cite{Feng2020} investigated the kinematic asymmetry of the ionized gas in a large sample of paired galaxies. In their study, the merging stage is determined by a combination of kinematic asymmetry ($\overline{v}_{\rm asym}$) and projected separation. The value of $\overline{v}_{\rm asym}$ is measured from \ha\ velocity maps of galaxies \citep[see details in][]{Feng2020}, which describes the asymmetry of velocity field contributed by the interaction-induced non-rotating motion. They find significantly enhanced star formation of paired galaxies with high kinematic asymmetries, while paired galaxies with low kinematic asymmetries show no significant enhancement of star formation rate (SFR) even at small projected separation. The enhancement of SFR is also tightly correlated with smaller projected separation. These results are consistent with previous findings \citep{Scudder2012,Patton2013}, which use only projected separation as the merging stage indicator. These findings suggest that the kinematic asymmetry is an effective indicator of galaxy mergers.

The enhancement of star formation during galaxy-galaxy interactions requires sustaining gas supply, which can be further explored by observations of cold gas (e.g., \hi\ gas ) in galaxy pairs. In this work, we compile a major-merger galaxy pairs sample selected from MaNGA survey to study the \hi\ content of merging galaxies, adopting the kinematic asymmetry and projected separation as indicators of the merging stage. We compare the \hi\ gas properties and star formation of the pair sample with a robustly matched control sample. Our study can provide better constraints on the \hi\ gas fraction in merging galaxies and, more importantly, study the HI gas depletion/replenishment in clearly defined interacting stages.

This paper is organized as follows. In Section \ref{sec:obs}, we introduce the sample selection of galaxy pairs, observation setup and data reduction of our work. We then present the main results of \hi\ gas properties and star formation in Section \ref{sec:results}, with a further discussion presented in Section \ref{sec:discussions}. Finally, we summarize the main results in Section \ref{sec:Summary}. Throughout the whole paper, we adopt the standard $\Lambda$CDM cosmology with $H_0 = 70$ km s$^{-1}$ Mpc$^{-1}$, $\Omega_M = 0.3$, and $\Omega_{\Lambda} = 0.7$.

\section{Samples, Observations and Data Reduction} \label{sec:obs}

\subsection{Pair Sample}\label{subsec:sample}
In our parent sample \citep{Feng2019,Feng2020}, the isolated galaxy pairs are selected following these criteria: (1) the projected separation for member galaxies: 5 $h^{-1}$ kpc $\leqslant d_{\rm p} \leqslant$  200 $h^{-1}$ kpc, (2) the line-of-sight velocity difference: $|\Delta v|\leqslant$ 500 km $\rm s^{-1}$, (3) each pair member only has one neighbor satisfying the above criteria, (4) at least one member galaxy of each pair has been observed in MaNGA survey \citep{Bundy2015}, and the member galaxy has more than 70\% spaxels with \ha\ emission at S/N $>$ 5 within 1.5 effective radius ($R_e$), and (5) we only study star-forming galaxies $(\rm{log}(\rm{sSFR/yr}^{-1})>-11)$ in this work. In order to study major-merger pairs, we constrain the mass ratio as $M_1/M_2< 3$, where $M_1$ and $M_2$ represent the stellar masses of primary galaxies and companions, respectively. After that, 243 sources are selected as major-merger pairs and kept in the pair sample. The \hi\ data used in this work are either obtained by our PI programs with FAST or extracted from HI-MaNGA survey \citep{Masters2019,Stark2021}. The final pair sample consists of 66 galaxy pairs with \hi\ detections at S/N $>$ 5. Details of sample observations and data reduction are described in the following sub-sections.   

\subsubsection{FAST Observations}
Through our FAST PI programs (PT2020\_0152, PI: Q.Z. Yu; PT2020\_0186, PI: T.T. Fang), we carried out \hi\ observations of 8 interacting galaxy pairs selected from the above sample in 2020 September. As a pilot survey, these 8 targets have been selected based on their  projected separations ($d_{\rm p}$) and kinematic asymmetry ($\overline{v}_{\rm asym}$). The kinematic asymmetry ($\overline{v}_{\rm asym}$) of each target is measured from the \ha\ velocity map through MaNGA data. Based on \cite{Feng2020}, the \ha\ velocity map of each source is divided into a sequence of concentric elliptical rings, fitted with the Fourier series: 
\begin{equation}
    V( a,\psi) =A_0(a)+\sum_{n=1}^N k_n (a)\cos[n(\psi-\phi_n(a))],
\end{equation}
where $\psi$ and $a$ represent the azimuthal angle in the galaxy plane and the semi-major axis of the ellipse, respectively. $A_0$ is the zero-order Fourier component, while $k_n$ and $\phi_n$ are the amplitude and the phase coefficient of the nth-order Fourier component, respectively. \cite{Feng2020} have used the ratio between high-order and first-order coefficients to describe the disturbing level of the velocity field. Considering the first-order coefficient $k_1$ measures the rotating motion caused symmetric pattern, the high-order coefficients ($k_2,\ k_3,\ k_4,\ k_5$) characterize the asymmetry of velocity field contributed by the non-rotating motion. The kinematic asymmetry at a given radius is calculated as
\begin{equation}
    v_{\rm asym}=\frac{k_2+k_3+k_4+k_5}{4k_1}.
\end{equation}
The kinematic asymmetry $\overline{v}_{\rm asym}$ for the entire galaxy is the average value of $v_{\rm asym}$ within 1 $R_e$. We use the $\overline{v}_{\rm asym}$ to measure the strength of tidal interactions for paired galaxies. Briefly, the larger $\overline{v}_{\rm asym}$ value represents a larger asymmetry of the velocity field, which indicates a stronger tidal interaction of the merging process. The $d_{\rm p}$ of our observing targets ranges from $\sim$ 7 $h^{-1}$ kpc to $\sim$ 80 $h^{-1}$ kpc, and $\overline{v}_{\rm asym}$ ranges from 0.024 to 0.193. 

The targeted galaxy pairs were observed with tracking mode in September, 2020, using the 19-beam L-band receiver of FAST, with a band coverage of 1.05-1.45 GHz, and a beam size of $\sim2.9$ arcmin \citep{Jiang2020}. Using a $\sim$ 1 s sampling time, the wide-band spectrometer (Spec(W)) can record both polarizations with a channel width of 7.629 kHz ($\sim$ 1.6 $\text{km s}^{-1}$) covering 65,536 channels. During observations for each target, the OFF-target positions were observed with outer beams for baseline subtraction. We set the OFF-target positions 5.8-11.5 arcminutes away from our targets based on the SDSS and the NRAO Very Large Array (VLA) Sky Survey \citep[NVSS;][]{Condon1998} results to avoid optical sources at similar redshift and radio continuum sources at 1.4 GHz. Flux calibration was achieved by injecting periodic noise into the receiving system during observations. We used low noise mode to reduce baseline ripples, and the median temperature of the noise diode is $\sim$ 1.1 K \citep{Jiang2020}. Each target was observed with the same configuration mentioned above, and in Column 2 of Table \ref{tab:emission}, we list the specific ON-target time of each source.

Data were reduced with {\tt astropy} \citep{Astropy2013} and {\tt scipy} \citep{SciPy2020}, following the standard procedure of the FAST extragalactic \hi\ survey pipeline \citep{WangZ2020}. We first applied the flux calibration with the injected noise signal to derive the antenna temperature of the spectrum \citep{Jiang2020}. After checking the consistency of polarization XX \& YY for each spectrum, we combined the two polarizations. The temperature of each spectra was converted to flux density, based on the gain factor of different beams on the receiver (e.g., $\sim$ 13.42-15.98 Jy/K at 1350 MHz, see Table 5 of \citealt{Jiang2020}). After flux calibration, we performed bandpass subtraction with calibrated ON-target and OFF-target spectra. Cubic spline fitting was applied to subtract the baseline of the spectrum. Radio frequency interference (RFI) was flagged manually during the subtraction process, and we exclude the RFI-contaminated data in later works of spectral line fitting and measurement.  For sources with multiple exposures, the spectra used for co-adding are weighted by $\text{1/rms}^2$ to get higher S/N, where the root mean square (rms) noise is measured from the signal-free region of the spectrum. The velocity of the final spectrum is Doppler-corrected and converted to the barycentric frame. In total, six galaxy pairs with \hi\ emission detections (S/N $>$ 5) are used for the following analysis.

\begin{figure*}[ht!]
\includegraphics[width=1\textwidth]{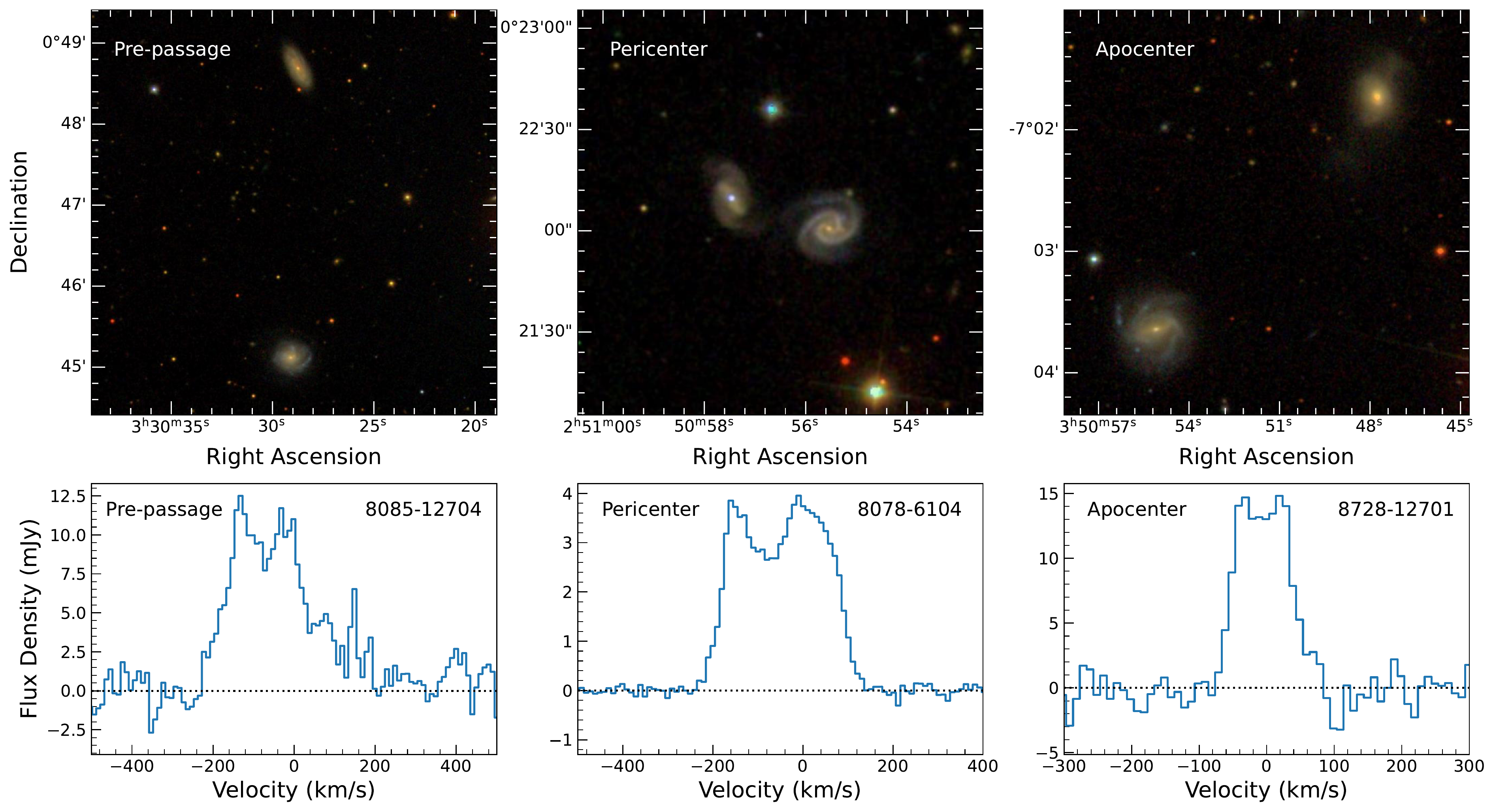}
\caption{The upper panels present examples of SDSS optical images of galaxy pairs at pre-passage, pericenter, and apocenter stages, respectively. The lower panels show the corresponding \hi\ line profiles of each galaxy pair. The MaNGA plate-ifu IDs are labeled on the top right of each spectrum.
\label{fig:opt}
}
\end{figure*}

\subsubsection{HI-MaNGA Survey}
The HI-MaNGA project mainly uses the Robert C. Byrd Green Bank Telescope (GBT) to  perform \hi\ follow-up observations for the SDSS-VI MaNGA survey, focusing  MaNGA-observed galaxies with stellar mass $8.5 <\text{log}(M_{\star}/M_{\odot})< 11.2$ and redshift $z<0.05$ \citep{Masters2019}. The newly released HI-MaNGA catalog \citep{Stark2021} contain 3669 galaxies with \hi\ and optical IFU data, including targets observed by GBT in 2018, and an updated crossmatch with the Arecibo Legacy Fast ALFA (ALFALFA) survey \citep{Haynes2018}. After a crossmatch between the HI-MaNGA second data release (DR2) catalog and the galaxy pair sample from \cite{Feng2020}, we obtain 187 paired galaxies with \hi\ emission line detections at S/N $>$ 5. Further constrained with the mass ratio $M_1/M_2< 3$ criteria, we select 60 major-merger pairs from HI-MaNGA survey. Together with 6 galaxy pairs observed with FAST, we adopt 66 major-merger pairs as the galaxy pair sample in the following analysis.

\subsection{Subsamples and Control Sample}\label{subsec:subsample}
Following \cite{Feng2020}, we divide the pair sample into two subsamples based on their $\overline{v}_{\rm asym}$ values: low asymmetry ($0.007<\overline{v}_{\rm asym}<0.029$), and high asymmetry ($0.029<\overline{v}_{\rm asym}<0.316$). We set the boundary of $\overline{v}_{\rm asym} =0.029$ to make sure at least 33 percent sources are included in the low asymmetry sample. Based on recent simulations, the velocity field of paired galaxy after the pericentric passage tends to be more disturbed than that before the passage \citep{Hung2016}. Therefore, we define the galaxy pairs with low $\overline{v}_{\rm asym}$ values as pairs at the stage of pre-passage, which means there are no significant interactions between paired galaxies. During the interactions between galaxy pairs, simulations suggest the physical separation constantly decreases before pericentric passage and increases when the paired galaxies are approaching apocenter \citep{Torrey2012,Moreno2015,Moreno2019}. In this work, we use the projected separation ($d_{\rm p}$) as a reference to define the pericenter and apocenter stages of merging galaxy pairs. The high $\overline{v}_{\rm asym}$ valued pairs with $d_{\rm p}<$ 50 $h^{-1}$ kpc are defined to be at the stage of pericenter passage, and the high $\overline{v}_{\rm asym}$ valued pairs with $d_{\rm p}>$ 50 $h^{-1}$ kpc are approaching the apocenter passage. In Figure \ref{fig:opt}, we present examples of the SDSS optical images and corresponding \hi\ line profiles of galaxy pairs at pre-passage, pericenter, and apocenter stages, respectively. The \hi\ spectra are provided either by the HI-MaNGA survey or our FAST observations. For each pair, both the component galaxies are covered inside the beam of the telescope. The number of pairs in each stage: (1) Pre-passage: 21 ; (2) Pericentric passage: 11; (3) Apocenter stage: 34.

We have compiled a control sample of isolated star-forming galaxies $\rm (log(sSFR/yr^{-1})>-11)$  to compare the effect of galaxy interactions on star formation and \hi\ gas properties. These isolated galaxies are also selected from MaNGA survey, defined as galaxies without any bright neighbors ($r < 17.77$). Neighbors are defined as companions satisfying $d_{\rm p} \leqslant$  200 $h^{-1}$ kpc and $|\Delta v|\leqslant$ 500 km $\rm s^{-1}$ \citep{Feng2020}. To avoid the bias towards galaxies with strong emission lines, control galaxies are also required to have 5$\sigma$ detection of \ha\ emission for at least 70\% spaxels within 1.5 $R_e$. By crossmatching with HI-MaNGA DR2 catalog, we selected 433 isolated galaxies with \hi\ detections at S/N $>$ 5 as our control sample pool. We performed a galaxy-by-galaxy matching between the paired galaxies and isolated galaxies from the control sample pool for further comparisons and analysis (see details in Section \ref{subsec:offset}). Although the systematic study of $\overline{v}_{\rm asym}$ for isolated galaxies is still ongoing, private communications with Feng did suggest a small number of isolated galaxies do have higher $\overline{v}_{\rm asym}$ values than those of pre-passage galaxy pairs, but the difference is not significant \citep{Feng2020}. In addition, the ongoing study of the $\overline{v}_{\rm asym}$ of isolated galaxies suggests there is a tight negative correlation between $\overline{v}_{\rm asym}$ and $M_{\star}$ for low mass isolated galaxies (log$(M_{\star}/M_{\odot})< 9.5$), while there is no correlation between $\overline{v}_{\rm asym}$ and the \hi\ gas fraction for isolated galaxies (Feng et al. in prep.). After carefully matching of $M_{\star}$, our results are not affected by the insignificant difference of the $\overline{v}_{\rm asym}$ between isolated galaxies and pre-passage pairs.

\startlongtable
\begin{deluxetable*}{lccccccccccc}

\tabletypesize{\footnotesize}
\tablenum{1}
\tablecaption{Galaxy Pairs Sample\label{tab:sources}}
\tablewidth{0pt}
\tablehead{
\colhead{MaNGA ID} & \colhead{R.A.} & \colhead{Dec.} & \colhead{$z$} & \colhead{$\text{log}(M_{\star})$} & \colhead{$\text{log(SFR)}$} & \colhead{$\overline{v}_{\text{asym}}$} & \colhead{$d_{\rm p}$} &  \colhead{log$(M_{\text{\hi}})$} & \colhead{Stage} & \colhead{Type} & \colhead{Source Blending}\\
\colhead{plate-ifu} & \colhead{(deg.)} & \colhead{(deg.)} & \colhead{} & \colhead{($M_{\odot}$)} & \colhead{($M_{\odot}\ \text{yr}^{-1}$)} & \colhead{} & \colhead{($h^{-1}$kpc)} & \colhead{($M_{\odot}$)} & \colhead{} & \colhead{} &\colhead{}
}
\decimalcolnumbers
\startdata
7495-6104	&	204.69273	&	26.49877	&	0.02950	&	9.43	&	$-$1.03	&	0.0272	&	155.30	&	9.26	&	1	&	S+S	&	N	\\
7968-3704	&	322.60787	&	$-$0.47452	&	0.01989	&	10.08	&	0.08	&	0.0629	&	31.75	&	9.78	&	2	&	S+S	&	Y	\\
8078-6104	&	42.73943	&	0.36941	&	0.04421	&	10.43	&	0.88	&	0.1930	&	18.35	&	9.88	&	2	&	S+S	&	Y	\\
8082-12703	&	49.51165	&	$-$0.53896	&	0.02104	&	10.48	&	0.18	&	0.0295	&	63.14	&	9.50	&	3	&	S+S	&	Y	\\
8085-12704	&	52.61990	&	0.81150	&	0.03080	&	10.72	&	0.37	&	0.0221	&	98.10	&	10.07	&	1	&	S+S	&	Y	\\
8134-12703	&	114.47365	&	47.95722	&	0.02097	&	9.35	&	$-$1.16	&	0.0220	&	155.10	&	9.43	&	1	&	S+E	&	N	\\
8137-9102	&	117.03863	&	43.59070	&	0.03114	&	10.57	&	0.16	&	0.0370	&	162.45	&	9.73	&	3	&	S+S	&	N	\\
8138-6104	&	117.29142	&	46.46188	&	0.03823	&	9.87	&	0.33	&	0.0250	&	97.11	&	10.11	&	1	&	S+S	&	Y	\\
8153-12701	&	39.63722	&	$-$0.86747	&	0.03923	&	10.26	&	0.03	&	0.0422	&	158.83	&	10.00	&	3	&	S+S	&	Y	\\
8241-6101	&	127.04853	&	17.37466	&	0.02087	&	8.82	&	$-$0.10	&	0.0761	&	31.83	&	9.52	&	2	&	S+S	&	Y	\\
8248-12703	&	136.74357	&	17.96378	&	0.02861	&	10.44	&	$-$0.64	&	0.2018	&	117.62	&	9.95	&	3	&	S+E	&	N	\\
8250-6101	&	138.75315	&	42.02439	&	0.02790	&	10.79	&	0.89	&	0.0396	&	44.97	&	9.88	&	2	&	S+S	&	Y	\\
8260-6101	&	182.40876	&	42.00967	&	0.02288	&	10.11	&	0.37	&	0.0293	&	28.93	&	9.50	&	2	&	S+S	&	Y	\\
8262-6101	&	184.66519	&	43.53872	&	0.02392	&	9.18	&	$-$0.40	&	0.0404	&	40.77	&	9.55	&	2	&	S+E	&	N	\\
8338-6102	&	172.68267	&	22.36354	&	0.02236	&	9.45	&	$-$0.09	&	0.0360	&	76.13	&	9.29	&	3	&	S+S	&	N	\\
8439-9102	&	143.75402	&	48.97674	&	0.02496	&	9.22	&	0.03	&	0.0463	&	176.08	&	9.43	&	3	&	S+S	&	N	\\
8450-6102	&	171.74883	&	21.14168	&	0.04177	&	10.17	&	0.67	&	0.0516	&	57.37	&	9.72	&	3	&	S+E	&	N	\\
8458-9102	&	146.97891	&	45.46848	&	0.01533	&	8.89	&	$-$0.93	&	0.0565	&	181.96	&	9.03	&	3	&	S+S	&	N	\\
8547-12702	&	217.91070	&	52.74946	&	0.04566	&	10.62	&	0.02	&	0.0187	&	137.28	&	10.40	&	1	&	S+E	&	N	\\
8548-1902	&	243.33996	&	48.15502	&	0.02007	&	9.10	&	$-$0.36	&	0.0214	&	177.94	&	9.64	&	1	&	S+S	&	N	\\
8552-12702	&	227.92840	&	43.97044	&	0.02758	&	9.50	&	0.14	&	0.0558	&	90.60	&	9.73	&	3	&	S+S	&	Y	\\
8552-6101	&	227.01700	&	42.81902	&	0.01801	&	8.66	&	$-$0.94	&	0.1178	&	100.06	&	9.04	&	3	&	S+E	&	N	\\
8566-12705	&	116.16133	&	40.36597	&	0.01982	&	9.97	&	$-$0.21	&	0.0479	&	55.61	&	10.04	&	3	&	S+S	&	Y	\\
8567-6102	&	119.31545	&	47.80304	&	0.02274	&	9.15	&	$-$0.55	&	0.0340	&	51.81	&	9.30	&	3	&	S+S	&	Y	\\
8588-12702	&	250.31305	&	39.29009	&	0.03054	&	9.71	&	$-$0.08	&	0.0648	&	88.57	&	9.95	&	3	&	S+S	&	N	\\
8611-12705	&	261.97903	&	59.72121	&	0.01809	&	9.17	&	$-$0.45	&	0.0770	&	106.01	&	9.44	&	3	&	S+S	&	Y	\\
8656-1901	&	7.71740	&	0.52876	&	0.01914	&	9.08	&	$-$0.31	&	0.0293	&	101.03	&	9.22	&	3	&	S+S	&	N	\\
8656-3703	&	7.75250	&	0.43575	&	0.01917	&	9.36	&	$-$0.55	&	0.0232	&	101.21	&	9.22	&	1	&	S+S	&	N	\\
8657-12704	&	10.40833	&	0.25764	&	0.01807	&	9.08	&	$-$0.50	&	0.0998	&	47.61	&	9.41	&	2	&	S+S	&	N	\\
8657-6104	&	10.41747	&	0.20879	&	0.01749	&	9.27	&	$-$0.32	&	0.0682	&	46.08	&	9.19	&	2	&	S+S	&	Y	\\
8713-12703	&	116.09855	&	40.34377	&	0.01956	&	9.97	&	$-$0.21	&	0.0483	&	55.01	&	10.02	&	3	&	S+S	&	Y	\\
8714-12702	&	118.68983	&	46.22018	&	0.02223	&	9.99	&	$-$0.29	&	0.0189	&	129.08	&	9.44	&	1	&	S+S	&	Y	\\
8716-12703	&	122.41118	&	54.55217	&	0.04210	&	9.62	&	0.04	&	0.0399	&	193.68	&	9.71	&	3	&	S+S	&	N	\\
8728-12701	&	57.72957	&	$-$7.06065	&	0.02841	&	10.50	&	$-$0.03	&	0.0716	&	66.78	&	9.78	&	3	&	S+E	&	N	\\
8942-12704	&	123.80750	&	27.40881	&	0.03767	&	9.41	&	$-$0.75	&	0.0522	&	148.79	&	9.60	&	3	&	S+S	&	N	\\
8977-12705	&	119.04436	&	33.24505	&	0.01730	&	9.05	&	$-$0.73	&	0.4318	&	56.75	&	9.27	&	3	&	S+S	&	Y	\\
8977-3704	&	118.77440	&	32.72867	&	0.01784	&	9.20	&	$-$0.13	&	0.0921	&	156.94	&	9.01	&	3	&	S+E	&	N	\\
8980-12704	&	225.59269	&	41.92164	&	0.01633	&	9.12	&	$-$0.48	&	0.0398	&	72.69	&	9.45	&	3	&	S+S	&	N	\\
8981-6101	&	185.94406	&	36.15281	&	0.03332	&	10.97	&	0.26	&	0.0320	&	81.77	&	9.89	&	3	&	S+S	&	Y	\\
8982-12703	&	203.80666	&	27.63004	&	0.02550	&	9.28	&	$-$0.92	&	0.0681	&	175.34	&	9.66	&	3	&	S+S	&	N	\\
8987-3701	&	136.24989	&	28.34772	&	0.04864	&	10.32	&	0.94	&	0.0157	&	44.75	&	10.08	&	1	&	S+S	&	Y	\\
8987-6101	&	137.23489	&	27.51382	&	0.02192	&	8.75	&	$-$1.09	&	0.0250	&	169.15	&	9.10	&	1	&	S+S	&	N	\\
8987-6103	&	137.26802	&	28.25682	&	0.02166	&	8.66	&	$-$1.85	&	0.1132	&	56.61	&	9.35	&	3	&	S+S	&	N	\\
8991-3703	&	176.02299	&	52.85804	&	0.01909	&	8.85	&	$-$0.95	&	0.0333	&	157.61	&	9.10	&	3	&	S+E	&	N	\\
8993-12704	&	165.44761	&	45.87374	&	0.02192	&	9.43	&	$-$0.22	&	0.2836	&	81.96	&	9.29	&	3	&	S+S	&	Y	\\
9027-12702	&	243.98368	&	31.40677	&	0.02256	&	10.00	&	0.06	&	0.0210	&	129.06	&	9.80	&	1	&	S+S	&	N	\\
9027-9101	&	243.90740	&	31.32130	&	0.02220	&	10.38	&	0.28	&	0.0352	&	126.97	&	9.89	&	3	&	S+S	&	Y	\\
9030-3702	&	241.23418	&	30.52578	&	0.05537	&	10.89	&	1.10	&	0.0219	&	48.35	&	10.49	&	1	&	S+S	&	Y	\\
9050-9101	&	245.99844	&	21.79503	&	0.03214	&	10.51	&	0.72	&	0.0310	&	41.92	&	9.50	&	2	&	S+S	&	Y	\\
9094-12705	&	240.46430	&	26.31944	&	0.04393	&	9.85	&	$-$0.11	&	0.0445	&	176.34	&	10.05	&	3	&	S+E	&	N	\\
9184-12703	&	118.58490	&	32.77462	&	0.01774	&	9.18	&	$-$0.49	&	0.0338	&	156.07	&	9.49	&	3	&	S+E	&	N	\\
9185-9101	&	256.21228	&	34.81733	&	0.05683	&	11.09	&	1.47	&	0.1930	&	7.25	&	10.09	&	2	&	S+S	&	Y	\\
9488-9102	&	126.70413	&	20.36485	&	0.02510	&	9.76	&	0.27	&	0.0804	&	190.82	&	10.31	&	3	&	S+S	&	N	\\
9492-12701	&	117.21679	&	17.35627	&	0.03955	&	10.59	&	0.64	&	0.0205	&	43.35	&	10.40	&	1	&	S+S	&	Y	\\
9493-12701	&	128.19586	&	22.57656	&	0.01540	&	9.42	&	$-$1.07	&	0.0143	&	118.28	&	9.44	&	1	&	S+E	&	N	\\
9499-12703	&	118.42323	&	26.49270	&	0.03742	&	10.82	&	0.07	&	0.0281	&	78.04	&	10.17	&	1	&	S+E	&	N	\\
9499-6102	&	119.94700	&	25.35696	&	0.02669	&	10.42	&	0.33	&	0.0290	&	159.52	&	9.38	&	3	&	S+S	&	N	\\
9505-9102	&	140.10889	&	27.49819	&	0.02682	&	9.80	&	$-$0.43	&	0.0232	&	47.33	&	9.53	&	1	&	S+S	&	Y	\\
9507-12701	&	128.25074	&	26.01405	&	0.01763	&	9.87	&	$-$0.47	&	0.0668	&	21.83	&	9.62	&	2	&	S+E	&	N	\\
9509-6103	&	123.06447	&	26.20257	&	0.02508	&	10.04	&	$-$0.06	&	0.0147	&	87.35	&	9.71	&	1	&	S+S	&	N	\\
9881-6102	&	205.21316	&	24.47331	&	0.02705	&	10.77	&	$-$0.05	&	0.0195	&	152.23	&	9.99	&	1	&	S+E	&	N	\\
9886-12704	&	237.86938	&	25.82014	&	0.02213	&	8.98	&	$-$1.02	&	0.0289	&	169.53	&	9.38	&	1	&	S+S	&	N	\\
9889-1902	&	234.85860	&	24.94357	&	0.02286	&	10.55	&	1.15	&	0.0240	&	7.92	&	9.43	&	1	&	S+S	&	Y	\\
9890-12705	&	233.52046	&	29.90688	&	0.03650	&	9.29	&	$-$0.31	&	0.0478	&	169.86	&	9.68	&	3	&	S+S	&	N	\\
10221-12703	&	124.05695	&	25.07600	&	0.01561	&	8.85	&	$-$0.78	&	0.0470	&	103.12	&	9.38	&	3	&	S+S	&	N	\\
10503-3703	&	160.02053	&	5.46299	&	0.02629	&	9.31	&	$-$0.70	&	0.0249	&	194.63	&	9.50	&	1	&	S+S	&	N	\\
\enddata
\tablecomments{The columns show (1) MaNGA ID of the targets; (2) R.A. in degree; (3) Dec. in degree; (4) redshift; (5) stellar mass from the MPA-JHU Catalog; (6) SFR from the MPA-JHU Catalog; (7) kinematic asymmetry; (8) projected speration in $h^{-1}$ kpc; (9) estimated \hi\ mass from HI-MaNGA DR2 and FAST observations; (10) merging stage based on $\overline{v}_{\text{asym}}$ and $d_{\rm p}$; 1, 2, and 3 represent pre-passage, pericenter and apocenter, respectively; (11) type of galaxy pair; (12) source blending in \hi\ observations, Y and N represent yes and no for source blending problems, respectively.}
\end{deluxetable*}

\subsection{Source Blending}\label{subsec:blend}

For \hi\ observations of galaxy pairs, blending caused by relatively low spatial resolution is the main challenge for single-dish radio telescopes. Because of the large beam size, resolving two components in close interacting pairs is often not possible for GBT, Arecibo, and FAST. Therefore, we treated each unresolved pair as a single source for the analysis of \hi\ content and other related galaxy properties. HI-MaNGA DR2 have considered the source confusion problem and calculated the likelihood distribution of \hi\ flux ratio ($R$) contributed by companions \citep{Stark2021}. Then the probability of $R>0.2$ ($P_{R>0.2}$), which means the companions contribute at least 20 $\%$ of total flux, is estimated to identify the confused sources. We adopt the threshold from \cite{Stark2021} and consider galaxies with $P_{R>0.2}>0.1$ as unresolved pairs in our sample.

For these unresolved galaxy pairs, we divide them into spiral-spiral (S+S) and spiral-elliptical (S+E) types through three methods: (1) by adopting the visual inspection results of Galaxy Zoo \citep{Lintott2008,Lintott2011}, we classify the component galaxies with $P_{\rm CS}\geqslant 0.6$ as spiral type, and the others as elliptical type, where $P_{\rm CS}$ is the debiased probability of the spiral type \citep[see][]{Lintott2011}; (2) we follow the method from \cite{Xu2010} and classify a galaxy with $u-r>2.22$ and $R_{50}/R_{90}<0.35$ as elliptical type, and the rest as spiral type; (3) we identify galaxies with a S\'ersic index \citep{Simard2011} of $n_S<2.5$ as spiral type \citep{Shen2003}, and the rest as elliptical type. We adopt the median of the aforementioned three results as the final classification. The final pair sample contains 51 S+S and 15 S+E pairs. There are 9 S+S and 2 S+E pairs in the pericentric passage. In contrast, there are 26 S+S and 8 S+E pairs in the apocenter stage. We considered these unresolved pairs as single sources when calculating related properties (e.g., $M_{\star}$, SFR, and \hi\ gas fraction). For S+S pairs, the $M_{\star}$ and SFR are sums of two sub-components, while only the spiral component is considered in an S+E pair. We adopt the global SFR and $M_{\star}$ from the public catalog of MPA-JHU DR7\footnote{\url{https://wwwmpa.mpa-garching.mpg.de/SDSS/DR7/\#derived}}.

Due to the large beam size, \hi\ observations of galaxy pairs with single-dish radio telescopes also suffer from the contamination of neighboring spiral galaxies \citep{Zuo2018}. For our FAST observations, we followed the method of \cite{Zuo2018} and carefully searched the neighboring spiral galaxies ($|\Delta v|\leqslant$ 500 km $\rm s^{-1}$) within 3$\arcmin$ from the pair center. We found no contaminating neighbors for each source. We also performed the same procedure to search the neighboring spiral galaxies ($|\Delta v|\leqslant$ 500 km $\rm s^{-1}$) with a searching radius of 10$\arcmin$ for the 60 pairs from HI-MaNGA. We found only one galaxy pair may have been contaminated by a neighbor outside the beam ($ d_{\rm p}\sim 451\ h^{-1}$ kpc, $|\Delta v|\sim$ 111 km $\rm s^{-1}$ relative to the pair), and we performed a correction based on the algorithm from \cite{Zuo2018}. Our selection of isolated galaxy pairs helps reduce the possibility of contamination from neighboring spiral galaxies.

In Table \ref{tab:sources}, we list the basic information and physical properties of the galaxy pair sample used in the following analysis. In column 1, we list the MaNGA plate-ifu IDs of each source. This ID shows that at least one component galaxy in the pair has MaNGA observation. The J2000 equatorial coordinates of each source are listed in columns 2 and 3. Column 4 lists the spectroscopic redshift from SDSS. In columns 5 and 6, we list the stellar mass and SFR taken from the MPA-JHU catalog. The kinematic asymmetry and projected separation measured from MaNGA data are listed in columns 7 and 8. Column 9 lists the \hi\ mass estimated either from HI-MaNGA or FAST data. In columns 10 and 11, we list the merger stage and the type of the galaxy pair. Column 12 records the source blending of each target.

\begin{deluxetable*}{lcccccccr}
\tablenum{2}
\tablecaption{Emission line results from FAST observations\label{tab:emission}}
\tablewidth{0pt}
\tablehead{
\colhead{MaNGA ID} & \colhead{$t_{\text{ON}}$} & \colhead{rms} & \colhead{$S_{\text{peak}}$} & \colhead{S/N} & \colhead{$S_{\text{int}}$} & \colhead{$W_{50}$} & \colhead{$W_{20}$} & \colhead{log $M_{\text{\hi}}$}\\
\colhead{plate-ifu} & \colhead{(min)} & \colhead{(mJy)} & \colhead{(mJy)} & \colhead{} & \colhead{($\text{mJy km s}^{-1}$)} & \colhead{($\text{km s}^{-1}$)} & \colhead{($\text{km s}^{-1}$)} & \colhead{($M_{\odot}$)}
}
\decimalcolnumbers
\startdata
8078-6104 & 72 & 0.23 & 3.95 & 39.5 & 936.39 $\pm$ 93.83 & 265.08 $\pm$ 4.53 & 319.06 $\pm$ 8.27 & 9.88 $\pm$ 0.06 \\
8260-6101 & 12 & 0.51 & 7.49 & 32.6 & 1310.67 $\pm$ 131.85 & 194.86 $\pm$ 5.34 & 233.89 $\pm$ 6.33 & 9.50 $\pm$ 0.07 \\
8981-6101 & 10 & 0.61 & 5.54 & 24.1 & 1519.05 $\pm$ 153.26 & 374.06 $\pm$ 8.35 & 424.31 $\pm$ 10.56 & 9.89 $\pm$ 0.06 \\
9050-9101 & 12 & 0.53 & 4.15 & 23.1 & 701.84 $\pm$ 71.95 & 186.57 $\pm$ 7.92 & 231.57 $\pm$ 8.88 & 9.50 $\pm$ 0.06 \\
9093-12701 & 13 & 0.44 & - & - & $<$ 660 & - & - & $<$ 9.91\\
9185-9101 & 13 & 0.74 & 4.35 & 13.6 & 827.73 $\pm$ 85.60 & 252.34 $\pm$ 16.87 & 467.91 $\pm$ 23.94 & 10.09 $\pm$ 0.06 \\
9889-1902 & 12 & 0.71 & 5.30 & 21.2 & 1115.48 $\pm$ 113.32 & 286.47 $\pm$	10.27 & 353.89 $\pm$ 17.46 & 9.43 $\pm$ 0.06 \\
10496-12704 & 13 & 0.70 & - & - & $<$ 1050 & - & - & $<$ 9.84 \\
\enddata
\tablecomments{The columns show (1) name of our targets; (2) total on source time; (3) noise level at $\sim$ 1.6 $\text{km s}^{-1}$; (4) peak flux density of the \hi\ emission line in mJy after smoothing to $\sim$ 10 $\text{km s}^{-1}$; (5) the peak signal-to-noise ratio measured after smoothing to $\sim$ 10 $\text{km s}^{-1}$; (6) integrated flux in $\text{mJy km s}^{-1}$ after smoothing to $\sim$ 10 $\text{km s}^{-1}$; (7) the \hi\ line width in $\text{km s}^{-1}$ measured at $50\%$ of the peak flux; (8) the \hi\ line width in $\text{km s}^{-1}$ measured at $20\%$ of the peak flux; (9) estimated \hi\ mass. All errors are calculated at 1$\sigma$ level.}
\end{deluxetable*}

\section{Results} \label{sec:results}

\subsection{ \texorpdfstring{\ion{H}{1}}{HI} Gas Properties}

\begin{figure*}[ht!]
\includegraphics[width=1\textwidth]{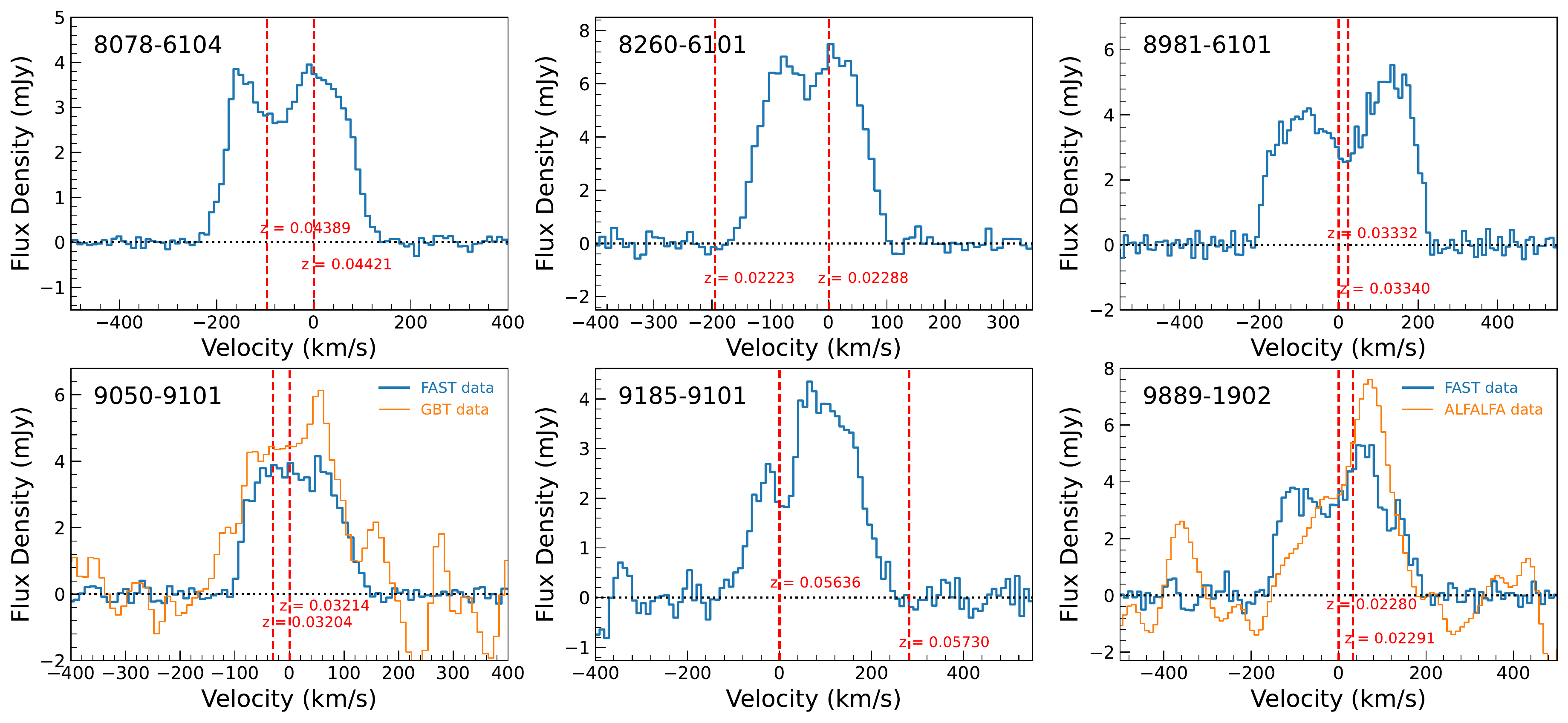}
\caption{\hi\ emission line profiles of merging galaxy pairs detected by FAST. The new FAST \hi\ spectra shown by blue lines have been calibrated, baseline-subtracted and smoothed to a velocity resolution of $\sim$ 10 $\text{km s}^{-1}$. The orange lines present \hi\ spectra from HI-MaNGA survey, which have been smoothed to a velocity resolution of $\sim$ 10 $\text{km s}^{-1}$.} The velocity of each spectrum is Doppler-corrected and converted to the barycentric frame, and the zero point is set based on the optical spectroscopy redshift of a member galaxy in each pair. The red dashed vertical lines show the velocities of each member galaxy.
\label{fig:emission}

\end{figure*}

We detect 6 \hi\ emission lines and 1 absorption line in FAST observations of 8 galaxy pairs. For the undetected sources, we assume a typical full width at half maximum (FWHM) of 300 $\text{km s}^{-1}$ and estimate the upper limit of the integrated flux as $5\times \text{rms}\times 300\ \text{mJy km s}^{-1}$. The detailed results of \hi\ emission lines are listed in Table \ref{tab:emission}, and the profiles of the emission line are shown in Figure \ref{fig:emission}. In Table \ref{tab:emission}, we list the source name and corresponding ON-source time in columns 1 and 2. The rms noise is $\sim$ 0.23 - 0.71 mJy measured at the velocity resolution of 1.6 $\text{km s}^{-1}$ and listed in column 3. We have measured the \hi\ peak flux density and integrated flux of each source after smoothing, which are listed in columns 4 and 6, respectively. In column 5, we list the peak signal-to-noise ratio. In columns 7 and 8, we measure the \hi\ velocity widths of the detected spectra at 50 and 20 percent level of the peak flux density ($W_{50}$ and $W_{20}$, see \citealt{Koribalski2004}). In the last column, we present the estimated \hi\ mass of each galaxy pair. 

In Figure \ref{fig:emission}, we present the new \hi\ line profiles observed by FAST. The spectra shown in blue lines are boxcar smoothed to a velocity resolution of $\sim$ 10 $\text{km s}^{-1}$. After smoothing, we calculate the \hi\ integrated flux density of each detection (Table \ref{tab:emission}). Among these 6 targets, 9050-9101 and 9889-1902 have previous \hi\ detections at S/N $>$ 5 in the HI-MaNGA survey, observed by GBT and Arecibo, respectively. To directly compare the line profile and flux, we overplot the \hi\ spectra from the HI-MaNGA survey in orange lines in Figure \ref{fig:emission}, and each spectrum has been boxcar and hanning smoothed to a velocity resolution of $\sim$ 10 $\text{km s}^{-1}$ \citep{Stark2021}. For target 9050-9101, we have measured the integrated flux from GBT data (954.64 $\text{mJy km s}^{-1}$), which is $\sim 26\%$ higher than that from FAST data (Table \ref{tab:emission}). For target 9889-1902, although the line profile is inconsistent (Figure \ref{fig:emission}), it does not affect the main result. We have measured the integrated flux of 9889-1902 from ALFALFA data (1126.31 $\text{mJy km s}^{-1}$), which is in good agreement with the FAST data (Table \ref{tab:emission}). 

We estimate the \hi\ mass using the following redshift-dependent formula \citep{Ellison2018,Stark2021}:
\begin{equation}
\frac{\mhi}{M_{\odot}} = \frac{2.356 \times 10^{5}}{(1+z)^2} \left(\frac{D_{\rm L}}{\text{Mpc}}\right)^2 \left(\frac{\int S(v) {\rm d}v}{\text{Jy km s}^{-1}}\right) , 
\label{eq:mhi}
\end{equation}
where the $D_{\rm L}$ is the luminosity distance of the targets in unit of Mpc, and the $S(v)$ is the Doppler-corrected \hi\ flux density in Jy, integrated with the Doppler velocity in $\text{km s}^{-1}$. The uncertainty is estimated by considering the rms noise measured from the line-free region and 10$\%$ systematic error of $\int S(v) {\rm d} v$. The systematic error is estimated based on measurements during observations, including uncertainty from flux calibration ($\sim$ 5-6$\%$), baseline subtraction, beam attenuation, and pointing.

For galaxy pairs in HI-MaNGA survey, we adopt the \hi\ mass from HI-MaNGA DR2 catalog, which is also calculated based on equation \ref{eq:mhi}. Adopting the derived \hi\ mass $(\mhi)$, we calculated the \hi\ gas fraction $(f_{\text{\hi}})$ and the star formation efficiency of \hi\ gas (\sfe). 

The \hi\ gas fraction $f_{\text{\hi}}$ is defined as:
\begin{equation}
f_{\text{\hi}} = \frac{\mhi}{M_{\star}}  . 
\label{eq:fhi}
\end{equation}
In some studies, the definition of $f_{\text{\hi}} = \mhi/(\mhi+M_{\star})$ is commonly used, but here we adopt the definition frequently used by previous studies \citep{Ellison2015,Ellison2018} for comparison. The mean \hi\ gas fraction of the pair sample in logarithm is $\text{log}\ f_{\text{\hi}}=-0.12 \pm 0.06$. In contrast, the mean \hi\ gas fraction of the control sample $(\text{log}\ f_{\text{\hi}}=-0.03 \pm 0.02)$ is higher than that of the pair sample, which indicates the control galaxies are more gas-rich on average. However, the $f_{\text{\hi}}$ is tightly correlated to the global stellar mass ($M_{\star}$) for star-forming galaxies \citep{Catinella2010}, so we perform a galaxy-by-galaxy match between the pair sample and control sample in Section \ref{subsec:offset}.

The \sfe\ is defined as:
\begin{equation}
{\rm SFE_{\text{\hi}}} = \frac{\rm{SFR}}{\mhi}  . 
\label{eq:sfe}
\end{equation}
We use \sfe\ to describe the star formation efficiency of \hi\ gas for galaxies in pairs and controls. The mean $\rm log\ (SFE_{\text{\hi}}/yr^{-1})$ values are $-9.80 \pm 0.06$ and $-9.96 \pm 0.02$ for galaxies in pairs and controls, respectively.

\subsection{Offset of Galaxy Properties}\label{subsec:offset}

In order to compare the \hi\ gas fraction and other properties of galaxies in pairs and controls along the merger sequence, we calculate the ``offset" quantities following the method of \cite{Ellison2015,Ellison2018}. Each paired galaxy is matched in stellar mass and redshift with at least five isolated galaxies in the control pool. We require the matched galaxies satisfying the tolerance of $|\Delta \text{log}\ (M_{\star}/M_{\odot})| < 0.2$ and $|\Delta z|< 0.01$ \citep{Feng2020}. If the minimum of matched controls is not reached, the tolerances of stellar mass and redshift are grown by 0.1 dex and 0.005, respectively. We can find enough matched controls for most of the paired galaxies in the first round.

We perform this galaxy-by-galaxy matching to reduce observational biases. By matching stellar mass, we mitigate the bias caused by the anti-correlation between the \hi\ gas fraction and the stellar mass, as revealed in previous studies \citep{Catinella2010,Cortese2011}. In the HI-MaNGA DR2, a part of \hi\ data come from the ALFALFA survey. We perform redshift matching to account for the strong distance dependence of \hi\ detection in the ALFALFA survey \citep{Haynes2018}. Besides, previous observations reveal the correlation between the \hi\ content of galaxies and the local environment, i.e. galaxies in cluster and group environments have lower \hi\ gas fractions \citep{Solane2001,Cortese2011,Hess2013}. However, our selection of samples requires the sources to be isolated in both the pair and the control samples (see Section \ref{subsec:sample}, \ref{subsec:subsample}), which reduces the bias due to environmental differences. Therefore, we do not perform additional environmental matching between the pair and the control samples. 

After matching of stellar mass and redshift, the offset of \hi\ gas fraction, for example, is calculated as 
\begin{equation}
    \Delta f_{\text{\hi}} = \text{log}\ f_{\text{\hi, pair}} - \text{log}\ \text{median}(f_{\text{\hi, control}}) ,
\end{equation}
where the $\text{log}\ f_{\text{\hi,pair}}$ represents the \hi\ gas fraction of paired galaxy, and the $\text{log}\ \text{median}(f_{\text{\hi,control}})$ is the median \hi\ gas fraction of its matched control galaxies in the logarithm scale. We apply the same procedure to compute the offsets $\Delta \text{SFR}$ and $\Delta \rm SFE_{\text{\hi}}$. Notice that offsets of these galaxy properties are calculated in the logarithm scale. The mean \hi\ gas fraction offset ($\Delta f_{\text{\hi}}$) of all galaxy pairs is $ -0.06\pm0.03$ dex, which indicates the $f_{\text{\hi}}$ of galaxies in pairs on average is $\sim$ 15\% deficient compared with isolated galaxies. The SFR offset of all galaxy pairs is $ \Delta \text{SFR} = 0.09\pm0.05$ dex, suggesting the SFR of paired galaxies and isolated galaxies on average have no significant difference. The marginally enhanced SFR ($\sim$ 23\%) is insignificant with a large uncertainty. The \sfe\ offset of the full pair sample is $ 0.12\pm 0.06$ dex, which means the \sfe\ in galaxy pairs on average is marginally enhanced $\sim$ 32 \% compared with control galaxies.

\begin{figure*}[ht!]
\includegraphics[width=1\textwidth]{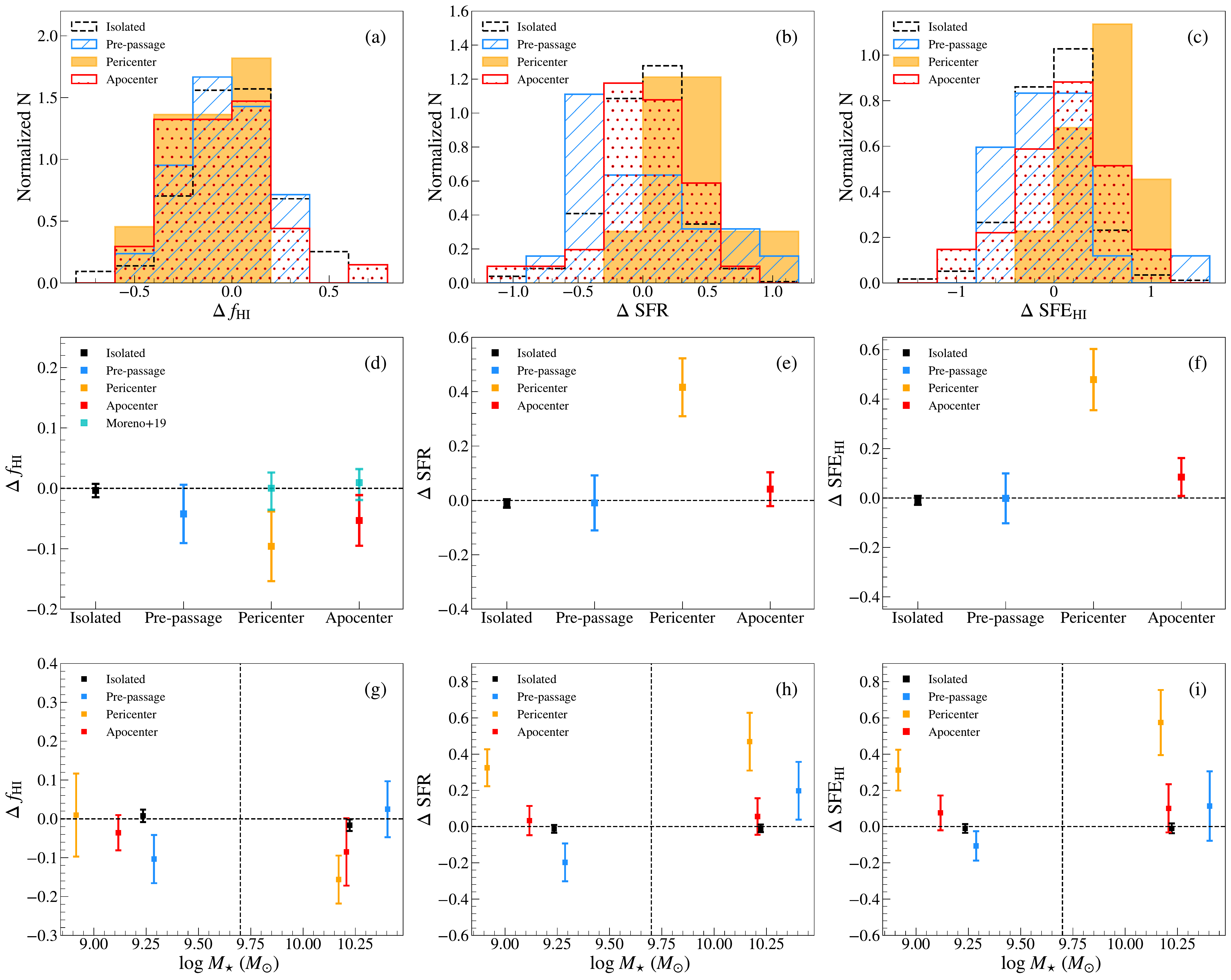}
\caption{The top panels show histograms of different galaxy property offsets. Paired galaxies at different merger stages and the control sample are marked with blue, orange, red and black, respectively. In each histogram, the number of bins is set to 8. (a) The distribution of $\Delta f_{\text{\hi}}$. (b) The distribution of $\Delta \text{SFR}$.  (c) The distribution of $\Delta \rm SFE_{\text{\hi}}$. The middle panels show the mean value distributions for different galaxy property offsets. Paired galaxies at different merger stages and the control sample are marked with blue, orange, red and black, respectively. In each plot, the mean value is indicated by square point in the middle and the error of the mean as error bar for each distribution. (d) The mean value of $\Delta f_{\text{\hi}}$ of our results and the simulation results (cyan data points) from \cite{Moreno2019}. (e) The mean value of $\Delta \text{SFR}$. (f) The mean value of $\Delta \rm SFE_{\text{\hi}}$. The lower panels present comparisons of galaxy properties of subsample in two $M_{\star}$ bins. Galaxy pairs at different merger stages are marked with blue, orange, and red, respectively. In each plot, the mean value is indicated by square point in the middle and the error of the mean as error bar. The vertical line in each plot indicates the boundary of $M_{\star}$ bins. (g) $\Delta f_{\text{\hi}}$ as a function of $M_{\star}$. (h) $\Delta \text{SFR}$ as a function of $M_{\star}$. (i) $\Delta \rm SFE_{\text{\hi}}$ as a function of $M_{\star}$.
\label{fig:offset}
}
\end{figure*}


We present our main results in Figure \ref{fig:offset}. The top panels show the histogram distributions of the offsets of different galaxy properties, and middle panels show the mean value distributions. Our results suggest mild \hi\ depletion occurs during merging, especially when pairs are at the pericenter stage. In Figure \ref{fig:offset} (a) and Figure \ref{fig:offset} (d), we plot the distributions of \hi\ gas fraction offsets for isolated galaxies and paired galaxies at different merger stages. As shown in Figure \ref{fig:offset} (a), we use black dashed bars to present the histogram of $\Delta f_{\text{\hi}}$ in the control sample, which indicates the intrinsic spread of the \hi\ gas fraction offset. The histograms of $\Delta f_{\text{\hi}}$ in pre-passage, pericentric passage, and apocenter passage are plotted with blue, orange, and red bars, respectively. In Figure \ref{fig:offset} (d), we present the mean values of $\Delta f_{\text{\hi}}$ for each subsample with black, blue, orange, and red squares, and the error bar represents the error of the mean value for each distribution. Compared to the controls, the mean values of each distribution indicate that on average the \hi\ gas fraction of pairs is marginally decreased at pericenter and apocenter stages for $ 0.10 \pm 0.05$ dex and $ 0.05 \pm 0.04$ dex ( $ \sim26\%$ and $ \sim12\%$, respectively). At the pre-passage stage, the \hi\ gas fraction of galaxy pairs ($  \Delta f_{\text{\hi}} = -0.04\pm0.05$ dex) is comparable to that of isolated galaxies. The cyan data points show the results of FIRE-2 simulation \citep{Moreno2019}. Their simulation predicts a $\sim 4\%$ enhancement of the cool gas mass ( mainly traced by \hi\ gas) on average during the galaxy-pair period. Comparing with their simulation results, our data indicate weak decreases of $f_{\text{\hi}}$ for pairs at pericenter and apocenter stages, respectively. 

The offset of SFR reveals the enhanced star formation of galaxy pairs encountering close interactions. In Figure \ref{fig:offset} (b) and Figure \ref{fig:offset} (e), we plot the distribution of global SFR offsets for isolated galaxies and paired galaxies at different merger stages. In Figure \ref{fig:offset} (b), the $\Delta \text{SFR}$ in pairs during pericentric passage tends to distribute at positive values, which indicates pairs at this stage present enhanced SFR. As shown in Figure \ref{fig:offset} (e), on average the paired galaxies during pericenter passage have strong SFR enhancement $ \Delta \text{SFR} = 0.42\pm0.11$ dex, and the mean $\Delta \text{SFR}$ values of paired galaxies during pre-passage and apocenter passage are $ -0.01 \pm 0.10$ dex and $ 0.04 \pm 0.06$ dex, respectively.

\begin{figure*}[ht!]
\includegraphics[width=1\textwidth]{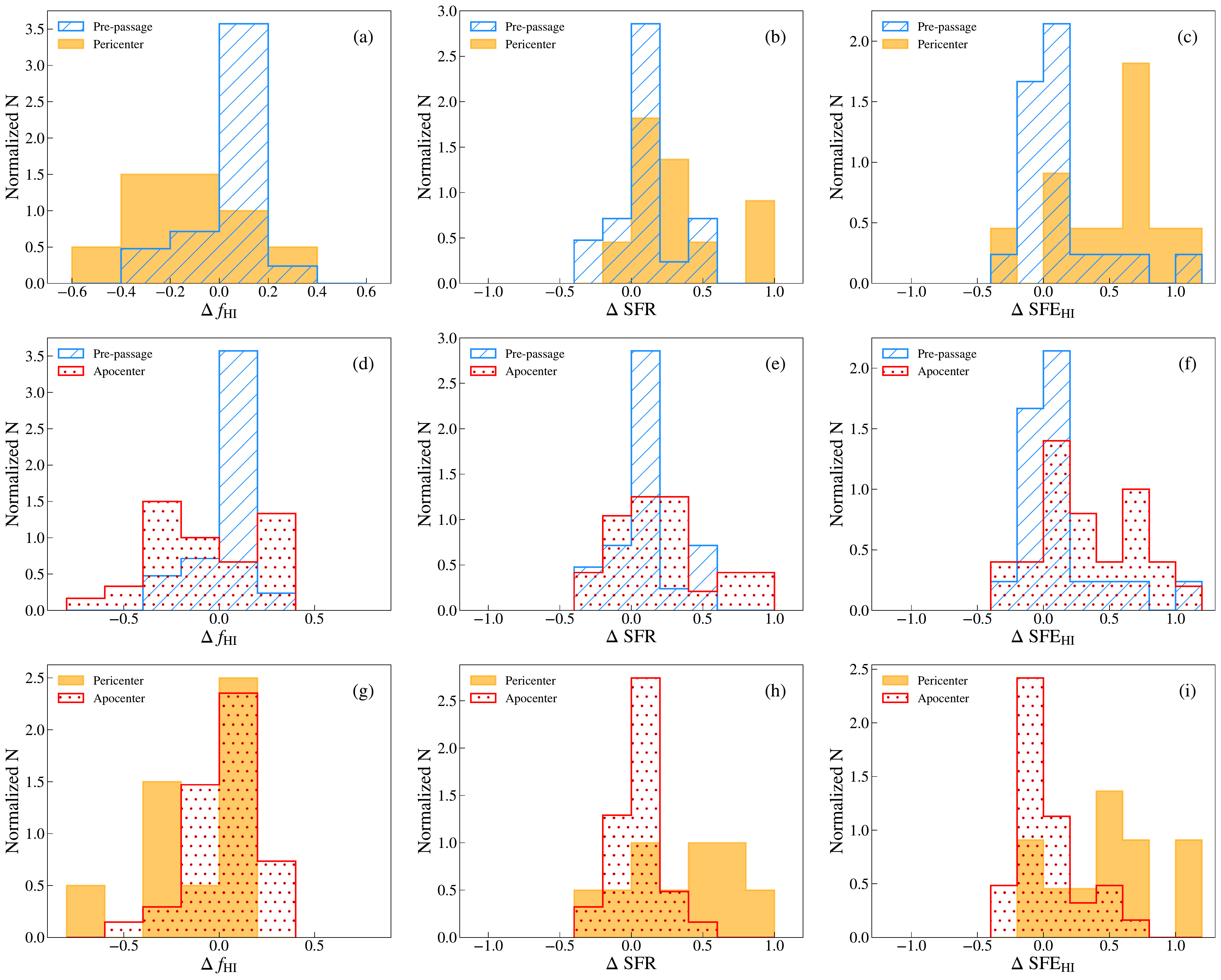}
\caption{Histograms of different galaxy property offsets between subsamples. Paired galaxies at different merger stages are marked with blue, orange, and red, respectively. (a) The distribution of $\Delta f_{\text{\hi}}$ when matching the pericenter sample with the pre-passage sample. (b) The distribution of $\Delta \text{SFR}$ when matching the pericenter sample with the pre-passage sample. (c) The distribution of $\Delta \rm SFE_{\text{\hi}}$ when matching the pericenter sample with the pre-passage sample. (d) The distribution of $\Delta f_{\text{\hi}}$ when matching the apocenter sample with the pre-passage sample. (e) The distribution of $\Delta \text{SFR}$ when matching the apocenter sample with the pre-passage sample. (f) The distribution of $\Delta \rm SFE_{\text{\hi}}$ when matching the apocenter sample with the pre-passage sample. (g) The distribution of $\Delta f_{\text{\hi}}$ when matching the pericenter sample with the apocenter sample. (h) The distribution of $\Delta \text{SFR}$ when matching the pericenter sample with the apocenter sample. (i) The distribution of $\Delta \rm SFE_{\text{\hi}}$ when matching the pericenter sample with the apocenter sample.  
\label{fig:subs}
}
\end{figure*}

Similar to SFR, our data suggest significantly enhanced \sfe\ of paired galaxies at the pericenter stage. In Figure \ref{fig:offset} (c) and Figure \ref{fig:offset} (f), we plot the distribution of \sfe\ offsets for isolated galaxies and paired galaxies at different merger stages. As shown in Figure \ref{fig:offset} (c), the distributions of $\Delta$\sfe\ for pairs at different stages are similar to these of $\Delta \text{SFR}$. Our results in Figure \ref{fig:offset} (f) indicate that the \sfe\ of paired galaxies during pericenter passage on average is significantly enhanced $ 0.48\pm 0.12$ dex, while the paired galaxies at pre-passage stage have mean the \sfe\ offset $\Delta \rm SFE_{\text{\hi}}=  0.00 \pm 0.10$ dex, which has no difference compared with the controls. The mean \sfe\ offset of pairs at apocenter stage is $\Delta \rm SFE_{\text{\hi}}= 0.08 \pm 0.08$ dex. Considering the \sfe\ of S+S pair is higher than that of S+E pairs as revealed by previous observations \citep{Zuo2018}, we further investigate whether the types of pairs caused the difference in \sfe\ between pericenter and apocenter stage. As mentioned in Section \ref{subsec:blend}, the ratios of S+E/S+S pairs are $ \sim 0.22$ and $ \sim 0.30$ for pairs at pericentric passage and apocenter stage, respectively. When excluding S+E pairs, the $\Delta \rm SFE_{\text{\hi}}$ values are $0.60\pm0.11$ and $0.08\pm0.08$ for pairs at pericentric passage and apocenter stage, respectively. These results suggest the \sfe\ difference is unlikely caused by the different types of pairs.

In the bottom panels of Figure \ref{fig:offset}, we present comparisons of galaxy properties of subsamples in two $M_{\star}$ bins. We consider galaxies with log$(M_{\star}/M_{\odot})<9.7$ as low mass galaxies while galaxies with log$(M_{\star}/M_{\odot})\geqslant 9.7$ as high mass galaxies in each subsample. As shown in Figure \ref{fig:offset} (g), the $f_{\text{\hi}}$ of paired galaxies at the pericenter stage and that of isolated galaxies have no significant difference in the low $M_{\star}$ bin. In the high $M_{\star}$ bin, pairs at the pericenter stage show decreased $f_{\text{\hi}}$ ($  \Delta f_{\text{\hi}} = -0.16\pm0.06$ dex) compared to isolated galaxies. The \hi\ gas fractions of the pre-passage sample and the apocenter sample have no significant difference from that of isolated galaxies. In Figure \ref{fig:offset} (h), paired galaxies at the pericenter stage show enhanced SFR ($ \Delta \text{SFR} = 0.32\pm0.10$ dex) relative to isolated galaxies in the low $M_{\star}$ bin, and the SFR enhancement ($ \Delta \text{SFR} = 0.47\pm0.16$ dex) is higher in the high $M_{\star}$ bin. In contrast, pairs at the apocenter stage have similar SFR in both $M_{\star}$ bins compared to isolated galaxies. For the pre-passage pairs, the SFR is marginally suppressed relative to isolated galaxies in the low $M_{\star}$ bin ($ \Delta \text{SFR} = -0.20\pm0.10$ dex), while the difference in SFR is not significant in the high $M_{\star}$ bin. The difference in \sfe\ is similar to that of SFR (Figure \ref{fig:offset} (i)). The \sfe\ offsets of pairs at the pericenter stage are $ 0.31 \pm 0.11$ dex and $ 0.57 \pm 0.18$ dex in the low and high $M_{\star}$ bins, respectively. Pre-passage pairs and apocenter pairs have similar \sfe\ in both $M_{\star}$ bins compared to isolated galaxies.

Since the numbers of galaxy pairs are different for each subsample, we further compare the aforementioned properties by matching the stellar mass and redshift of subsamples. As shown in Figure \ref{fig:subs}, we adopt the same matching method and calculation of offsets for the analysis of subsamples. Because of the limited galaxies in subsamples, a couple of pairs have only 1 or 2 matched controls. When matching the pericentric pairs with pre-passage pairs, our results suggest that on average the $f_{\text{\hi}}$ of pairs at the pericenter stage is $0.20\pm0.09$ dex lower than that of pre-passage pairs (Figure \ref{fig:subs} (a)). As shown in Figures \ref{fig:subs} (b) and (c), the SFR and \sfe\ of pericentric pairs are enhanced by $0.33\pm0.10$ dex and $0.51\pm0.12$ dex relative to the pre-passage pairs. When matching the apocenter pairs with the pre-passage pairs as controls, our results in Figure \ref{fig:subs} (d) indicate the $f_{\text{\hi}}$ of pairs at the apocenter stage is similar to pre-passage pairs ($ \Delta f_{\text{\hi}} = -0.02\pm0.06$ dex). In Figures \ref{fig:subs} (e) and (f), our data suggest $ \Delta \text{SFR} = 0.04\pm0.10$ dex and $\Delta \rm SFE_{\text{\hi}}=   0.05 \pm 0.11$ dex. When matching the pericentric pairs with the apocenter pairs as controls, we find the $f_{\text{\hi}}$ of pairs at the pericenter stage is $ 0.10\pm0.10$ dex lower than that of apocenter stage pairs (Figure \ref{fig:subs} (g)). The SFR and \sfe\ of pericentric pairs are enhanced by $ 0.42\pm0.14$ dex and $ 0.48\pm0.12$ dex compared to the apocenter pairs (Figure \ref{fig:subs} (h) and (i) ).

\section{Discussions} \label{sec:discussions}

\subsection{ \texorpdfstring{\ion{H}{1}}{HI} Depletion and SFR Enhancement at Different Merger Stages}
In this section, we discuss the \hi\ depletion and star formation enhancement of pairs at different merger stages, based on the offset of $f_{\text{\hi}}$, SFR and \sfe\ (Figure \ref{fig:offset}). Through the galaxy-by-galaxy matching of stellar mass and redshift, we have calculated the offset of $f_{\text{\hi}}$, SFR and \sfe\ for each subsample. When the pairs are at the pre-passage stage, there are no significant interactions between member galaxies. Therefore, the \hi\ gas fraction, SFR, and \sfe\ of pairs are basically similar to that of isolated galaxies, which has been presented by our results (Figure \ref{fig:offset}). 

During the pericentric passage, our analysis suggest that the SFR of member galaxies is significantly enhanced (a factor of $ \sim 2.6$), which is consistent with previous studies of close galaxy pairs \citep{Patton2013}. As revealed in previous CO observations of close pairs, the molecular gas fraction is also significantly enhanced and correlated to the enhancement of SFR \citep{Pan2018,Violino2018,Lisenfeld2019}. The enhancement of molecular gas fraction may originate from an accelerated transition from atomic to molecular gas caused by external pressure, which can arise in the early stage of the merger \citep{Kaneko2017}. In this scenario, the moderate decrease of $f_{\text{\hi}}$ revealed by our work can be explained as \hi\ gas depletion in fuelling the $\rm H_2$ reservoir. Furthermore, the significantly enhanced \sfe\ indicated by our data may because of the enhanced molecular-to-atomic gas mass ratio in close interacting pairs \citep{Lisenfeld2019}, considering the \sfe\ $=\text{SFE}\times(M_{\rm H_2}/\mhi)$, where the SFE is not enhanced compared with isolated galaxies \citep{Casasola04,Lisenfeld2019}. Alternatively, some recent observations of close pairs found small ($<$ a factor of 2) SFE enhancement \citep{Pan2018,Violino2018}, which may also drive the enhanced \sfe\ revealed by our data.

\begin{deluxetable*}{lcccc}
\tablenum{3}
\tablecaption{Spearman's rank order coefficient and the significance\label{tab:corre}}
\tablewidth{0pt}
\tablehead{
\colhead{Sample} & \colhead{$\Delta f_{\text{\hi}}$ vs. log $\overline{v}_\text{asym}$} & \colhead{$\Delta \text{SFR}$ vs. $\Delta f_{\text{\hi}}$} & \colhead{$\Delta \text{SFR}$ vs. log $\overline{v}_\text{asym}$} & \colhead{$\Delta \rm SFE_{\text{\hi}}$ vs. log $\overline{v}_\text{asym}$}
}
\decimalcolnumbers
\startdata
Pre-passage & $-$0.21 (0.37) & 0.28 (0.21) & $-$0.12 (0.62) & $-$0.06 (0.79) \\
Pericenter & 0.13 (0.71) & $-$0.32 (0.34) & 0.17 (0.61) & 0.06 (0.85) \\
Apocenter & 0.23 (0.19) & $-$0.11 (0.55) & $-$0.19 (0.29) & $-$0.31 (0.07) \\
All pairs & 0.03 (0.82) & $-$0.03 (0.81) & 0.09 (0.48) & 0.08 (0.53) \\
Isolated galaxies & 0.07 (0.14) & 0.08 (0.08) & 0.07 (0.15) & 0.02 (0.71)
\enddata
\tablecomments{This table lists Spearman's rank order coefficient $r_s$ with p-value in the parathesis as the significance. The columns are: (1) sub-samples of pairs at different stages and the whole pairs sample; (2) the Spearman's rank order coefficient and significance of $\Delta f_{\text{\hi}}$ vs. log $\overline{v}_\text{asym}$; (3) the Spearman's rank order coefficient and significance of $\Delta \text{SFR}$ vs. $\Delta f_{\text{\hi}}$; (4) the Spearman's rank order coefficient and significance of $\Delta \text{SFR}$ vs. log $\overline{v}_\text{asym}$; (5) the Spearman's rank order coefficient and significance of $\Delta \rm SFE_{\text{\hi}}$ vs. log $\overline{v}_\text{asym}$.}
\end{deluxetable*}

At the apocenter stage, our data show the mean \hi\ gas fraction of pairs remains suppressed compared to isolated galaxies. Compared to pairs during the pericentric passage, the \hi\ gas reservoir seems to be mildly replenished. This could be the cooling of hot/warm gas from CGM \citep{Moster2011,Tonnesen2012}, as the interaction-induced shocks at the pericenter stage can be gradually alleviated when pairs are approaching the apocenter. As for SFR, the previously enhanced star formation of pairs during pericentric passage is decreased at the apocenter stage. The decrease of SFR enhancement is in agreement with the previous result that galaxy pairs with large projected separations present weak SFR enhancement \citep{Scudder2012,Patton2013}. Simulations also suggest the merger-induced SFR enhancement is gradually decreased when the paired galaxies are approaching the apocenter after the first pericentric passage \citep{Moreno2015,Moreno2019}. As discussed above, the change of \sfe\ can be driven by SFE and $M_{\rm H_2}/\mhi$, but previous CO studies lack observations of pairs at the apocenter stage. Therefore, further CO observations of our pair sample will be helpful to determine the dominating factor for the observed suppression of \sfe\ enhancement.

\subsection{Correlation Analysis}
To further investigate correlations between $\Delta f_{\text{\hi}}$, $\Delta \text{SFR}$, $\Delta \rm SFE_{\text{\hi}}$, and $\overline{v}_\text{asym}$, we have performed Spearman's rank order analysis among these quantities. The results are listed in Table \ref{tab:corre}, including the full pair sample and subsamples of pairs at pre-passage, pericenter, and apocenter stages. Briefly, we describe the goodness of a correlation with Spearman's rank order coefficient ($r_s$; by definition, $-1 \leqslant r_s \leqslant  1$)  and consider $|r_s| \geqslant 0.6$ or $0.3 < |r_s| < 0.6$ as a tight or weak correlation, and $|r_s| \leqslant 0.3$ as no correlation. The p-value of Spearman's rank order describes the significance of each correlation. The significance (p-value) is the possibility of the assumption that the null hypothesis (no correlation) is correct.

As shown in Table \ref{tab:corre}, our results suggest no correlations between these properties of the full pair sample. Furthermore, we compare correlations between these properties with pairs at different stages. The coefficient $r_s$ of our data suggests no significant correlation between $\Delta f_{\text{\hi}}$ and log $\overline{v}_\text{asym}$ for pairs at different stages. We find there is no correlation between $\Delta f_{\text{\hi}}$ and $\Delta \text{SFR}$ for the full pair sample, which is consistent with the results of \cite{Ellison2018}. Among subsamples, only pairs at the pericenter stage present a insignificant negative correlation of $\Delta \text{SFR}$ vs. $\Delta f_{\text{\hi}}$ $ (r_s=-0.32, p=0.34)$. $\Delta \text{SFR}$ of pairs at different stages all show no correlation with log $\overline{v}_\text{asym}$ (Table \ref{tab:corre}). In terms of $\Delta \rm SFE_{\text{\hi}}$ vs. $\overline{v}_\text{asym}$,  only pairs at pericenter stage show a weakly negative correlation $ (r_s=-0.31, p=0.07)$. All these correlation analysis for subsamples are based on limited number of galaxies and need to be confirmed with larger samples. In addition, we include the control sample in this correlation analysis for comparison. Our results suggest no correlations between these properties for isolated galaxies (Table \ref{tab:corre}).


\begin{figure*}[ht!]
\includegraphics[width=1\textwidth]{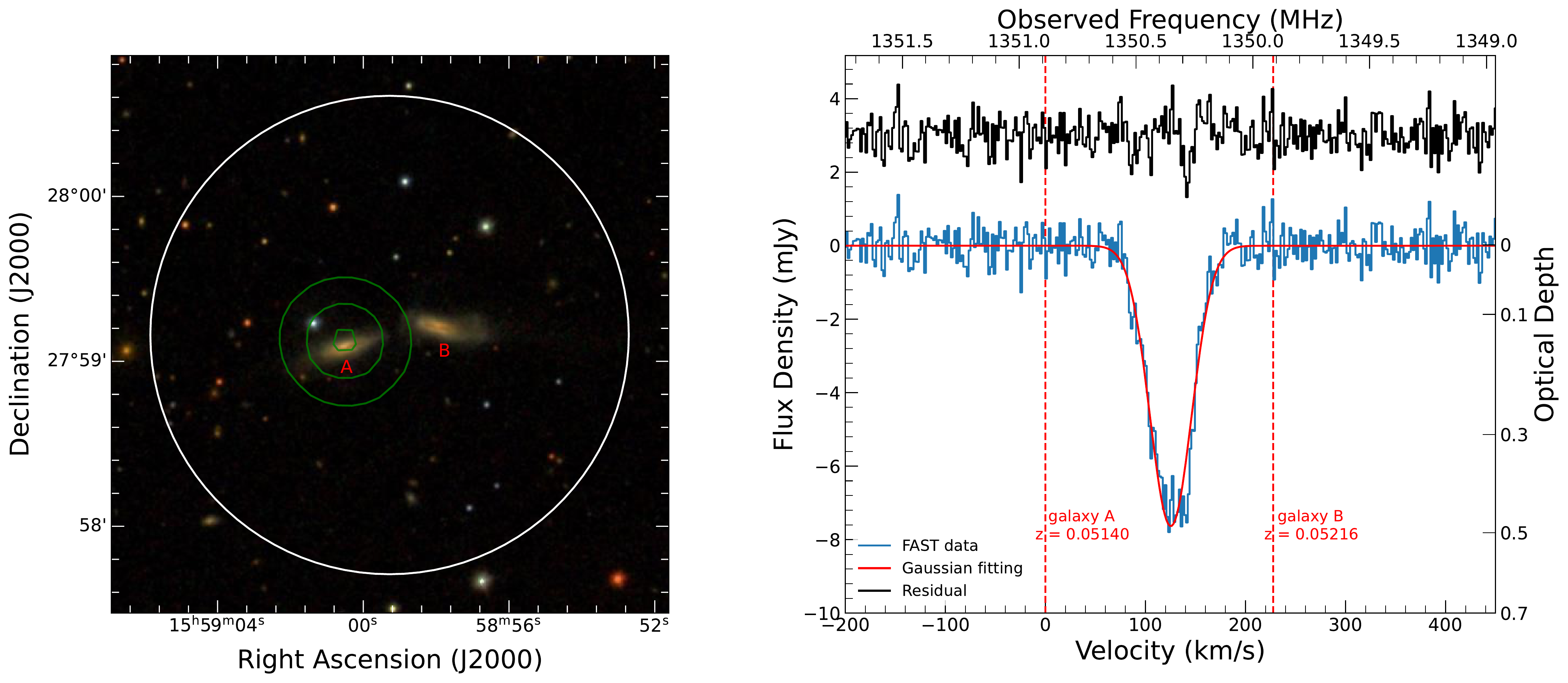}
\caption{Left: SDSS false color image of the merging galaxy pair 9093-12701/9094-12703 with NVSS radio continuum contours (green; 0.9, 14, 17 mJy) at 1.4 GHz. The white circle represents the beam ($\sim 2.9\arcmin$) of FAST. Right: The \hi\ 21 cm absorption spectrum of the merging galaxy pair at the velocity resolution $\sim$ \textbf{1.6} $\text{km s}^{-1}$. The blue line marks the observational data from FAST, and the red line shows a single Gaussian component fit to the spectrum. The black line on the top present the fitting residual spectrum with an +3 mJy offset for clarity. The velocity of the spectrum is corrected to the optical redshift of galaxy A at the barycentric frame, and the redshifts of the two galaxies are shown in red dashed lines. The optical depth on the right is calculated assuming a 20 mJy continuum source and the ${T_s}$ = 100 K, $C_f$ = 1 (see the text for details).
\label{fig:fig3}}
\end{figure*}

\subsection{Comparison with Previous Studies} \label{subsec:comparison}
Previously, \cite{Zuo2018} find no significant difference in the \hi\ gas fraction between merging and control galaxies, based on the \hi\ data of 70 galaxy pairs. Their pair sample has a mean \hi\ gas fraction of $\text{log}\ f_{\text{\hi}}=-1.12\pm0.06$, which is consistent with the results obtained using a large sample of $\sim$1000 star-forming galaxies with $10 <\rm{log}(M_{\star}/M_{\odot})< 11.5$ \citep[$\text{log}\ f_{\text{\hi}}\sim -1$,][]{Catinella2010}. In contrast, the mean \hi\ gas fraction of our pair sample is $\text{log}\ f_{\text{\hi}}= - 0.13\pm0.06$, which indicates the paired galaxies in our sample are more \hi-rich than those of \cite{Zuo2018} by a factor of $\sim$ 10. Considering the tight anti-correlation between $f_{\text{\hi}}$ and $M_{\star}$, we calculated the mean $\text{log}\ f_{\text{\hi}}$ of pairs with $\rm{log}(M_{\star}/M_{\odot})> 10$ for a comparison, and the result $(\text{log}\ f_{\text{\hi}}=-0.64\pm0.06)$ still show that our sample is more \hi-rich than that of \cite{Zuo2018} by a factor of $\sim$ 3. The control sample used in this work has a mean $\text{log}\ f_{\text{\hi}}=-0.03\pm0.02$, which presents a higher \hi\ gas fraction than that of the pair sample. When comparing with more massive galaxies ($\rm{log}(M_{\star}/M_{\odot})> 10$), the control sample shows a mean $\text{log}\ f_{\text{\hi}}= - 0.49\pm0.02$, which is higher than the results of \cite{Catinella2010} by a factor of $\sim$ 3.

Due to selection effects, both our pair and control samples have higher \hi\ gas fractions. We first investigate the effect due to the selection of \hi-detected (S/N $>$ 5) sources. We adopt the Kaplan-Meier estimator \citep{Feigelson1985} to re-calculate the mean $\text{log}\ f_{\text{\hi}}$ for the pair and control samples including non-detections (S/N $<$ 5). Including non-detections, the mean \hi\ gas fractions of the pair and control samples are $\text{log}\ f_{\text{\hi}}=-0.70\pm0.07$ and $\text{log}\ f_{\text{\hi}}=-0.94\pm0.06$, respectively. Indeed, the selection of \hi-detected (S/N $>$ 5) sources biases up the \hi\ fractions of both the pair and the control samples. As clarified in Section \ref{subsec:offset}, these direct comparisons can also be significantly affected by the mismatch of stellar mass and redshift between different samples. Besides, we require the galaxies in both our pair and control samples to have strong \ha\ emission for the measurement of kinematic asymmetry. This may also result in the selection of more gas-rich galaxies. 

Although our pair sample might be biased towards gas-rich galaxies, the selection of appropriate control sample can provide reasonable comparisons between our work and previous studies. By performing galaxy-by-galaxy matching, \cite{Ellison2018} found the mean \hi\ gas fraction of post-mergers is enhanced by a factor of $\sim$ 3 ($\Delta f_{\text{\hi}} = 0.51 $ dex) compared with isolated galaxies in xGASS \citep{Catinella2018}. Excluding \hi\ non-detections of the post-merger sample, the enhancement of mean \hi\ gas fraction ($\Delta f_{\text{\hi}} = 0.14$ dex) is still significant \citep{Ellison2018}. As shown in Section \ref{subsec:offset}, our results indicate the \hi\ gas fraction of galaxies in pairs, on average, is decreased $ \sim 15\%$ ($ \Delta f_{\text{\hi}} = -0.06\pm0.03$ dex) compared with the control sample selected in HI-MaNGA. Although the decrease of $f_{\text{\hi}}$ is marginal (at $ \sim 2\sigma$ level), our data suggest the \hi\ gas in major-merger pairs is depleted instead of enhanced during the galaxy-pair period. In contrast, the observed enhancement of $f_{\text{\hi}}$ in post-mergers may mainly originate from the arithmetic combination of the \hi\ gas in pre-merger pairs \citep{Ellison2018}. In addition, the progenitor galaxies of the post-mergers may contain minor-merger companions which have been already gas-rich.

Recent CO observations show that the molecular gas reservoir is enriched in interacting galaxy pairs \citep{Pan2018,Lisenfeld2019}. Considering the enhancement of molecular gas fraction requires fuelling from the \hi\ gas reservoir, the observed decrease of $f_{\text{\hi}}$ is in agreement with previous CO studies of close galaxy pairs.


\subsection{A New \texorpdfstring{\ion{H}{1}}{HI} Absorber Discovered by FAST}

We discovered a \hi\ 21-cm absorber in a close galaxy pair. This pair consists of two edge-on spiral galaxies with a projected separation $\sim$ 24.4 $h^{-1}$ kpc undergoing major-merger interactions. The member galaxies of this pair were observed as 9093-12701/9094-12703 (hereafter A and B, respectively) in MaNGA survey. The two galaxies shown in Figure \ref{fig:fig3} are identified as SDSS J155900.65+275907.4 (galaxy A) and SDSS J155858.08+275914.7 (galaxy B), and the redshifts based on SDSS spectra \citep{Abolfathi2018} of galaxy A \& B are 0.05140 $\pm$ 0.00001 and 0.05216 $\pm$ 0.00001, respectively. The previous optical survey \citep{Veilleux1995} classified the galaxy A nucleus as Seyfert 2 AGN, while the center of galaxy B is classified as a Low-ionization nuclear emission-line region (LINER). As in radio observations, the only continuum source of this system detected by the NVSS locates at the position of galaxy A (see the contour on the left panel in Figure \ref{fig:fig3}), and the flux density is 20 $\pm$ 1 mJy at 1.4 GHz \citep{Condon1998}. Considering there is only one continuum source in the beam of FAST, we confirm the background source of our detected \hi\ absorption is the AGN of galaxy A. We did not detect the \hi\ emission line of this pair because the continuum flux is probably higher than the emission line flux. Nevertheless, we estimated the up-limits for integrated \hi\ flux and the corresponding \hi\ mass with the rms noise (Table \ref{tab:emission}). 


The detected \hi\ 21 cm absorption line is shown in Figure \ref{fig:fig3}, with the zero point of velocity corrected to the optical redshift of galaxy A, which is most likely the continuum source of the absorption feature. The redshift of absorption is consistent with the merging system, indicating that the \hi\ absorbing gas is intrinsic and may originate from the environment of the merging pair. We performed a single-component Gaussian fit to the line profile of absorption feature, suggesting a relatively narrow line with FWHM = 49.09 $\pm$ 0.90 $\text{km s}^{-1}$ redshifted to the AGN at $v$ = 125.34 $\pm$ 0.38 $\text{km s}^{-1}$. A 20 mJy continuum flux density at 1.4 GHz \citep{Condon1998} is adopted to calculate the optical depth, which yields a peaked optical depth $\tau_{\rm peak}$ = 0.49 $\pm$ 0.05 and the integrated optical depth $\tau_{\rm int}$ = 14.50 $\pm$ 0.32 $\text{km s}^{-1}$. The column density of \hi\ absorbing gas $\nhi$ is estimated with the following equation:
\begin{equation}
\frac{\nhi}{\text{ cm}^{-2}} = 1.823 \times 10^{18} \frac{T_s}{c_f} \int \tau {\rm d} v ,  
\label{eq:2}
\end{equation}
where ${T_s}$ presents the spin temperature of \hi\ gas and $c_f$ is the source covering factor. Assuming ${T_s}$ = 100 K, $c_f$ = 1, the estimated column density for this \hi\ absorber $\nhi$ = $4.73 \pm 0.10 \times 10^{21} \text{ cm}^{-2}$, which is relatively high and consistent with the typical $\nhi$ value of \hi\ absorbers associated with mergers \citep{Dutta2018,Dutta2019}. The \hi\ absorber is impressive as the background continuum source is very faint while the \hi\ colunm density is relatively high. The continuum flux density from galaxy A ($S_{1.4}= 20\pm 1$ mJy) is among the weakest source with \hi\ 21 cm absorption detected in the local universe. FAST achieved the detection at S/N $\sim$ 17.3 with a short on-source time (780 s), showing outstanding sensitivity.

The velocity shift between our detected \hi\ absorbing gas and the merging system can give us indications on the origin of the absorbing gas. As shown in the right panel of Figure \ref{fig:fig3}, the \hi\ absorber is redshifted $\sim 125.3$ $\text{km s}^{-1}$ with respect to galaxy A, suggesting the atomic gas infall towards galaxy A. One possible scenario to explain the origin of the absorbing gas is that the gas is driven by the merging process and infalling toward the AGN of galaxy A, which could probably be responsible for triggering and fuelling the central AGN. Such inflow evidence has been observed in a merging system by \cite{Srianand2015}. Alternatively, the absorbing gas may originate from the rotating gas disk of the host galaxy \citep{Gallimore1999}. The alignment of the continuum from AGN and the foreground dense \hi\ gas disk can also result in the absorption line we detected. Both of the two possible origins cannot be determined or ruled out by only using the results of our observations because the beam size of FAST is big, and we cannot resolve the distribution, morphology, and kinematics of the absorbing gas. Therefore, high spatial resolution \hi\ observations of this source in the future will be helpful to untangle this problem.



\section{Summary} \label{sec:Summary}
In this paper, we investigate the \hi\ content and star formation of 66 major-merger galaxy pairs selected from MaNGA survey. We define merger stages of galaxy pairs with the combination of kinematic asymmetry and projected separation. The \hi\ properties are obtained from observations with FAST and archival data from HI-MaNGA DR2 \citep{Stark2021}. With \hi\ observations with FAST, we report the detection and properties of 6 \hi\ 21 cm emission lines and 1 absorption line. We compare the \hi\ properties and star formation of the pair sample with the control sample, and our findings are as follows.

1. Our data suggest the \hi\ gas fraction of major-merger galaxy pairs is marginally decreased compared with the isolated galaxies, indicating mild \hi\ gas depletion during merging. Through the galaxy-by-galaxy matching of stellar mass and redshift with the control sample, we calculate offsets of $f_{\text{\hi}}$, SFR, and \sfe\ for our pair sample. On average, the $f_{\text{\hi}}$ of galaxies in pairs is lower than that of the isolated galaxies by $ 0.06\pm0.03$ dex.

2. Our results indicate that galaxy pairs during pre-passage have similar $f_{\text{\hi}}$, SFR, and \sfe\ compared with isolated galaxies, while pairs at the pericenter stage have moderately decreased $f_{\text{\hi}}$ and significantly elevated SFR, \sfe. Pairs approaching the apocenter have insignificant \hi\ deficiency and insignificant SFR, \sfe\ enhancement. The mean $\Delta f_{\text{\hi}}$ values of pairs at pre-passage, pericenter, and apocenter stages are $ -0.04\pm0.05$ dex, $ -0.10\pm0.05$ dex, and $ -0.05\pm0.04$ dex, respectively. The mean $\Delta \text{SFR}$ values of pairs at pre-passage, pericenter and apocenter stages are $ -0.01\pm0.10$ dex, $ 0.42\pm0.11$ dex, and $ 0.04\pm0.06$ dex, respectively. The mean $\Delta \rm SFE_{\text{\hi}}$ values of pairs at pre-passage, pericenter, and apocenter stages are $ 0.01\pm0.10$ dex, $ 0.48\pm0.12$ dex, and $ 0.08\pm0.08$ dex, respectively. 

3. We find no significant correlations of $\Delta f_{\text{\hi}}$ vs. log $\overline{v}_\text{asym}$, $\Delta \text{SFR}$ vs. log $\overline{v}_\text{asym}$, $\Delta \text{SFR}$ vs. $\Delta f_{\text{\hi}}$ ,and $\Delta \rm SFE_{\text{\hi}}$ vs. log $\overline{v}_\text{asym}$ for the full pair sample. The Spearman's rank order coefficient suggests pairs at the pericenter stage show a insignificant negative correlation ($ r_s=-0.32,\ p=0.34$) of $\Delta \text{SFR}$ vs. $\Delta f_{\text{\hi}}$. Pairs at the apocenter stage present a weakly negative correlation ($ r_s=-0.31,\ p=0.07$) of $\Delta \rm SFE_{\text{\hi}}$ vs. log $\overline{v}_\text{asym}$.


4. The \hi\ 21 cm absorber we detected has a relatively high column density $\nhi$ = $4.73 \pm 0.10 \times 10^{21} \text{ cm}^{-2}$. The velocity shift indicates the absorbing gas may be infalling toward the central AGN. Determining the origin of the absorbing gas will be helpful to study the triggering of AGN in the merging process.

We conclude that our study marginally detects the \hi\ gas depletion during the galaxy-pair period of the merger. This is consistent with previous observations \citep{Hibbard1996,Georgakakis2000,Pan2018,Lisenfeld2019}. The suppressed \hi\ gas fraction may originate from the gas consumption in fuelling the enhanced $\rm H_2$ reservoir of galaxy pairs. It is a new method that we use kinematic asymmetry as the indicator of galaxy interaction. We can further continue this study as more and more IFU data become available. Our results are based on the global properties of galaxies in pairs. Thus, spatially resolved \hi\ observations will help us further explore the interplay between galaxy interactions and the \hi\ gas. Combining observations of the molecular gas for our sample, we will be able to build a complete picture of the cold gas evolution during mergers in the local universe.   

\acknowledgments
We thank the anonymous referee for the very helpful comments and suggestions that improved the paper. Q.Z.Y appreciates helpful discussions about the FAST data reduction with Cheng Cheng. This work is supported by the National Key R\&D Program of China No. 2017YFA0402600, and National Natural Science Foundation of China (NSFC) grants No. 11890692, 12133008, 12103017, and 11903056. We acknowledge the science research grants from the China Manned Space Project with NO. CMS-CSST-2021-A04. S.F. acknowledges support from Natural Science Foundation of Hebei Province (No. A2021205001), and Science Foundation of Hebei Normal University (No. L2021B08). FAST is operated and managed by the National Astronomical Observatories, Chinese Academy of Sciences. 

This work makes use data of the HI-MaNGA survey, based on observations performed with the Green Bank Observatory and the ALFALFA survey. The Green Bank Observatory is a facility of the National Science Foundation operated under cooperative agreement by Associated Universities, Inc. Data of the ALFALFA survey are based on observations with the Arecibo Observatory. The Arecibo Observatory is operated by SRI International under a cooperative agreement with the National Science Foundation (AST-1100968), and in alliance with Ana G. M\'endez-Universidad Metropolitana, and the Universities Space Research Association.

Funding for the Sloan Digital Sky Survey IV has been provided by the Alfred P. Sloan Foundation, the U.S. Department of Energy Office of Science, and the Participating Institutions. SDSS-IV acknowledges support and resources from the Center for High-Performance Computing at the University of Utah. The SDSS website is \url{www.sdss.org}.

SDSS-IV is managed by the Astrophysical Research Consortium for the Participating Institutions of the SDSS Collaboration including the Brazilian Participation Group, the Carnegie Institution for Science, Carnegie Mellon University, the Chilean Participation Group, the French Participation Group, Harvard-Smithsonian Center for Astrophysics, Instituto de Astrof{\'i}sica de Canarias, The Johns Hopkins University, Kavli Institute for the Physics and Mathematics of the Universe (IPMU)/University of Tokyo, the Korean Participation Group, Lawrence Berkeley National Laboratory, Leibniz Institut f{\"u}r Astrophysik Potsdam (AIP), Max-Planck-Institut f{\"u}r Astronomie (MPIA Heidelberg), Max-Planck-Institut f{\"u}r Astrophysik (MPA Garching), Max-Planck-Institut f{\"u}r Extraterrestrische Physik (MPE), National Astronomical Observatories of China, New Mexico State University, New York University, University of Notre Dame, Observat{\'a}rio Nacional/MCTI, The Ohio State University, Pennsylvania State University, Shanghai Astronomical Observatory, United Kingdom Participation Group, Universidad Nacional Aut{\'o}noma de M{\'e}xico, University of Arizona, University of Colorado Boulder, University of Oxford, University of Portsmouth, University of Utah, University of Virginia, University of Washington, University of Wisconsin, Vanderbilt University, and Yale University.
%



\software{Astropy \citep{Astropy2013}, Scipy \citep{SciPy2020}, LMFIT \citep{lmfit2014}, ASURV \citep{Feigelson1985}
}



\bibliography{ref}{}
\bibliographystyle{aasjournal}


\end{CJK*}
\end{document}